\documentclass[12pt]{article}
 
\usepackage{amssymb,amsmath}
 
\usepackage{accents} 

\usepackage{dsfont}

\usepackage{color}

\usepackage{xcolor}
\definecolor{ForestGreen}{RGB}{34,139,34}
\definecolor{mauve}{rgb}{0.7,0,0.43}
\definecolor{dkgreen}{rgb}{0,0.6,0}
\definecolor{darkgreen}{rgb}{0,0.6,0}
\definecolor{darkorange}{rgb}{1.0, 0.55, 0.0}
\definecolor{lightblue}{rgb}{0,0.2,0.5}
\definecolor{blue1}{rgb}{0,0.1,0.9}

\definecolor{lightblue}{rgb}{0,0.2,0.5}

\usepackage{graphics} 

\usepackage{subcaption}

\usepackage{float} 
\usepackage{pgf} 

\usepackage{siunitx}
\usepackage[colorlinks=true, urlcolor=blue,linkcolor=blue, citecolor=lightblue]{hyperref}

\usepackage{textcomp}

\usepackage{accsupp}    

\newcommand{\noncopynumber}[1]{
    \BeginAccSupp{method=escape,ActualText={}}
    #1
    \EndAccSupp{}
}
  	
\usepackage{listings}
\lstdefinelanguage{Maple}{
    morekeywords={proc, if, return, map, op, int, do, local, nops, convert, end, Iterator, SetPartitions, option, remember, Dimension, ToSets, Multiply, MatrixExponential, list},
    sensitive=false, 
    morecomment=[l]{//}, 
    morecomment=[s]{/*}{*/}, 
    morestring=[b]" 
} 

\DeclareMathAlphabet{\eufrak}{U}{}{}{} 
\SetMathAlphabet\eufrak{normal}{U}{euf}{m}{n}
\SetMathAlphabet\eufrak{bold}{U}{euf}{b}{n}

\usepackage{color}
\definecolor{lightblue}{rgb}{0,0.2,0.5}

\usepackage{tikz}
\usepackage{tkz-berge}
\usetikzlibrary{arrows,automata}
  
\numberwithin{equation}{section}

\newenvironment{Proof}{\removelastskip\par\medskip
\noindent{\em Proof.} \rm}{\penalty-20\null\hfill$\square$\par\medbreak}

\newenvironment{Proofy}{\removelastskip\par\medskip
\noindent{\em Proof} \rm}{\penalty-20\null\hfill$\square$\par\medbreak}

\allowdisplaybreaks

 \def\real{{\mathord{\mathbb R}}}
 
 \def\inte{{\mathord{\mathbb N}}}
 
 \def\qu{{\mathord{\mathbb Z}}}

 \def\real{{\mathord{{\rm I\kern-3pt R}}}}        
 
 \def\inte{{\mathord{{\rm I\kern-3pt N}}}}
 \def\sZZ{{\rm Z\kern-.45em{}Z}}

 \def\sQQ{{\kern 0.27em \vrule height1.45ex width0.03em depth0em
           \kern-0.30em \rm Q}}
 \def\qu{{\mathchoice
         {\sQQ}
         {\sQQ}
   {\kern 0.225em \vrule height1.05ex width0.025em depth0em \kern-0.25em \rm Q}
   {\kern 0.180em \vrule height0.78ex width0.020em depth0em \kern-0.20em \rm Q}
         }}
 \def\sGG{{\kern 0.27em \vrule height1.45ex width0.03em depth0em
           \kern-0.30em \rm G}}
 \def\gg{{\mathchoice
         {\sGG}
         {\sGG}
   {\kern 0.225em \vrule height1.05ex width0.025em depth0em \kern-0.25em \rm G}
   {\kern 0.180em \vrule height0.78ex width0.020em depth0em \kern-0.20em \rm G}
         }}

 \newtheorem{prop}{Proposition}[section]
 \newtheorem{lemma}[prop]{Lemma}
 
 \newtheorem{corollary}[prop]{Corollary}

\def\E{\mathop{\hbox{\rm I\kern-0.20em E}}\nolimits}

\usepackage[T1]{fontenc}
\usepackage{textcomp}

\DeclareMathSymbol{\mlq}{\mathord}{operators}{``}
\DeclareMathSymbol{\mrq}{\mathord}{operators}{`'}

\usepackage{enumerate} 

\usepackage{chngcntr}

\newcounter{hyp}
\newcommand{\X}{\Bbb X}            
\newcommand{\re}{\mathrm{e}}            

\title{
 \huge
 Closed-form modeling of neuronal spike train statistics 
 using multivariate Hawkes cumulants 
} 
 
\author{
\large 
 Nicolas Privault\thanks{nprivault@ntu.edu.sg} 
\\ 
\small
Division of Mathematical Sciences 
\\ 
\small 
School of Physical and Mathematical Sciences 
\\ 
\small
Nanyang Technological University 
\\ 
\small 
21 Nanyang Link 
\\ 
\small
Singapore 637371
}

\author{
\large 
 Nicolas Privault\footnote{
\href{mailto:nprivault@ntu.edu.sg}{nprivault@ntu.edu.sg}
}
\\ 
\small 
Division of Mathematical Sciences 
\\ 
\small
Nanyang Technological University 
\\ 
\small 
21 Nanyang Link, Singapore 637371 
\and
Mich\`ele Thieullen\footnote{
\href{mailto:michele.thieullen@sorbonne-universite.fr}{michele.thieullen@sorbonne-universite.fr}
} 
\\ 
\small ~~  LPSM-UMR 8001 - Case Courrier 158 
\\ 
\small ~~ Sorbonne Université 
\\ 
\small ~~ 4 Place Jussieu, 75252 Paris Cedex 05,
 France 
}

\usepackage{etoolbox}
\let\bbordermatrix\bordermatrix
\patchcmd{\bbordermatrix}{8.75}{4.75}{}{}
\patchcmd{\bbordermatrix}{\left(}{\left[}{}{}
\patchcmd{\bbordermatrix}{\right)}{\right]}{}{}

\usepackage[shortcuts]{extdash} 

\usepackage{scalerel}

\begin{document}

\hyphenation{func-tio-nals} 
\hyphenation{Privault} 

\maketitle 

\vspace{-0.9cm}

\baselineskip0.6cm
 
\begin{abstract} 
 We derive exact analytical expressions for the cumulants of any orders 
 of neuronal membrane potentials driven by spike trains 
 in a multivariate Hawkes process model with excitation and inhibition.
 Such expressions can be used for the prediction and sensitivity analysis
 of the statistical behavior of the model over time,
 and to estimate the probability densities of neuronal membrane potentials
 using Gram-Charlier expansions. 
 Our results are shown to provide a better alternative to Monte Carlo estimates
 via stochastic simulations,
 and computer codes based on combinatorial recursions are included.
\end{abstract} 
 
\noindent {\bf Key words:} 
Multivariate Hawkes processes;
filtered shot noise processes; 
multivariate cumulants;
Gram-Charlier expansions;
excitatory synapses;
inhibitory synapses;
membrane potentials. 
 
\baselineskip0.65cm
 
\section{Introduction}
Hawkes processes \cite{hawkes1971} are self-exciting point processes
that have been applied to the modeling of random spike trains in neuroscience
in e.g. \cite{cardanobile},
\cite{krumin},
\cite{gerhard},
\cite{ychen}. 
Neuronal spike train activity has been modeled using multivariate
Hawkes processes in e.g. \cite{reynaud-bouret2}, \cite{ocker},
\cite{kordovan}, where filtered Hawkes processes have been interpreted
as free membrane potentials in the linear-nonlinear cascade model. 
In this framework, the cumulants of
multivariate Hawkes processes yield important statistical
information. 
However, the analysis of statistical properties of Hawkes processes is made
difficult by their recursive nature,
 in particular, computing the cumulants of Hawkes processes
involves technical difficulties due to the infinite recursions involved.

\medskip

Neuronal synaptic input has also been modeled using multiplicative Poisson shot noise
 driven by random current spikes, in e.g. \cite{verveen}, \cite{tuckwell},
 see also 
\cite{kuhn},
\cite{rudolph},
\cite{richardson},
\cite{burkitt},
 for the analysis of stationary limits
 in the case of constant Poisson arrival rates,
 and \cite{wolff,wolff2}, see also 
 \cite{amemori}, \cite{burkitt2}, \cite{mclaughlin}
 for time-dependent Poisson intensities
 modeling of time-inhomogeneous synaptic input.
 In this framework, the time evolution of the 
 probability density functions of membrane potentials has been
 described in \cite{brigham-destexhe}, \cite{neuron} 
 by Gram-Charlier probability density expansions based on moment and cumulant estimates. 

\medskip

The computation of the moments of Hawkes processes has been the object
of several approaches, see \cite{dassios-zhao2}, \cite{cui} and \cite{daw} 
 for the use of differential equations, 
 and \cite{bacry} for 
 stochastic calculus methods applied to 
 first and second order moments.
Other techniques have been introduced for linear and nonlinear self-exciting processes,
including Feynman diagrams \cite{ocker}, path integrals \cite{kordovan},
and tree-based methods \cite{jovanovic} applied up to third order cumulants. 
However, such methods appear difficult to implement systematically for higher order
cumulants, and they use finite order expansions that only
approximate cumulants even in the linear case. 

\medskip 

In this paper, we provide a recursion for the closed-form computation of the
cumulants of multivariate Hawkes processes, without involving approximations.
 For this, we extend the recursive algorithm of \cite{hawkescumulants} 
to the computation of joint cumulants of all orders
of multivariate Hawkes processes. 
 This algorithm, based on a recursive relation for the Probability Generating Function (PGFl) of self exciting point processes started from a single point, relies on sums over partitions and Bell polynomials. 
 In what follows, we will apply this algorithm to 
 Hawkes processes with inhibition, by using 
 negative weights in their cluster point process construction.
 We note that although our cumulant expressions are
 proved only for non-negative weights,
 the results remain numerically accurate and consistent with the
 sampled cumulants of Hawkes processes with inhibition 
 as long as the process does not become inactive over long time intervals,
 see also \S~1 of \cite{ocker}.
 
\medskip 
 
 In Proposition~\ref{p1} and Corollary~\ref{c1} we compute the joint cumulants 
 of membrane potentials modeled according to a filtered Hawkes process
 as in \cite{ocker}. 
 In comparison with Monte Carlo simulation estimates, explicit expressions 
 allow for immediate numerical evaluations over multiple ranges of parameters,
 whereas Monte Carlo estimations can be slow to implement. 
 In addition, such expressions are
 suitable for algebraic manipulations and tabulation, 
 e.g. they can be differentiated in closed form with respect to time to
 yield the dynamics of cumulants, or with respect to
 any system parameter to yield sensitivity measures.
 Numerical applications of our closed form expressions
 are presented in Section~\ref{s3}, where they are compared to
 Monte Carlo estimates. 
 Although our simulations in Figures~\ref{fig4.11-0} to \ref{fig4.22-0} have been run with 10 million samples,
 Monte Carlo estimates of higher-order cumulants 
 can be subject 
 to numerical instabilities not observed with closed-form expressions.
 In particular, they become degraded starting with 
 joint third cumulants (see Figure~\ref{fig400-3}-$b)$) and
 fourth cumulants (see Figure~\ref{fig4.22-0}-$a)$),
 and they become clearly insufficient for the estimation of
 fourth joint cumulants (see Figure~\ref{fig4.22-0}-$b)$).  

 \medskip 
  
 Closed-form cumulant expressions are then applied in Section~\ref{s5} to the
 explicit derivation of cumulant-based Gram-Charlier expansions
 for the probability density function of the
 membrane potentials at any given time. 
 showing that densities are negatively skewed with positive excess kurtosis. 

 \medskip 

 We proceed as follows. 
 In Section~\ref{s2} we present closed-form recursions for the
 computation of cumulants of any order in a multivariate Hawkes process
 model. 
 Numerical results are then presented in Section~\ref{s3}
 with application to the modeling of connectivity in spike
 train statistics. 
 In Section~\ref{s5} we present
 numerical experiments based on 
 cumulants for the estimation of probability densities of potentials
 by Gram-Charlier expansions. 
 In the appendices we present the derivation of recursive cumulant
 and moment identities 
 for the closed-form computation of the moments of Hawkes processes,
 in the multivariate case, 
 with the corresponding codes written in Maple and Mathematica. 

\section{Cumulants of multivariate Hawkes processes} 
\label{s2}
 This section describes our algorithm for the computation of cumulants.
 Let $(H_1(t),\ldots , H_n (t))_{t\geq 0}$ denote a multivariate linear Hawkes
 point process 
 with self-exciting stochastic intensities of the form 
 \begin{equation}
   \label{fjlkf1} 
 \lambda_i(t) := \nu_i(t) + \sum_{j=1}^n \int_0^t \gamma_{i,j} ( t-s) d H_j(s), \qquad t\in \real_+, 
\end{equation} 
 with Poisson offspring intensities $\gamma_{i,j}(dx)=\gamma_{i,j}(x)dx$ and
 possibly time inhomogeneous Poisson
 baseline intensities $\nu_i(dt) = \nu_i(t)dt$, $i=1,\ldots , m$.
 The next proposition provides a way to compute the joint 
cumulants of random sums 
 by an induction relation based on the Bell polynomials. 
 In what follows, we assume that $\gamma_1 ( \real_+ ) + \cdots + \gamma_m (\real_+ ) < 1$, and 
 consider the integral operator $\Gamma$ defined as 
 $$
 ( \Gamma f ) (x,i) = \sum_{j=1}^m \int_0^\infty f(x+y ,j) \gamma_{i,j} (dy), \qquad
 x\in \real_+, \ i=1,\ldots , m, 
$$
 and, letting $I$ denote identity, the inverse operator $(I-\Gamma)^{-1}$ given by 
\begin{eqnarray*} 
\lefteqn{
 ( (I-\Gamma)^{-1} f ) (x ,i ) 
 = 
 f(x,i) + \sum_{n=1}^\infty ( \Gamma^n f ) (x,i)
}
  \\
   & = & 
  f(x,i) + \sum_{n=1}^\infty
 \sum_{j_1,\ldots , j_n=1}^m
 \int_0^\infty \cdots \int_0^\infty f(x+y_1+\cdots + y_n,j_n)
 \gamma_{i,j_1} ( dy_1) \cdots \gamma_{j_{n-1},j_n} ( dy_n), 
\end{eqnarray*} 
 $x\in \real_+$, $i=1,\ldots , m$. 
The following statements hold for the joint cumulants 
 $\kappa_{(x,i)}^{(n)}(f_1,\ldots , f_n)$  
of $ 
\left(
\sum_{j=1}^m \int_0^\infty f_i(t,j) dH_j(t),
\ldots  ,
\sum_{j=1}^m \int_0^\infty f_n(t,j) dH_j(t)
\right)$
 given that the multidimensional Hawkes process 
 is started from a single jump located in
 $H_i(t)$ at time $x\in \real_+$, $i=1,\ldots , m$. 
\begin{prop}
\label{p1}
\begin{enumerate}[a)]
  \item The first cumulant 
    $\kappa_{(x,i)}^{(1)}(f)$ of $ 
    \sum_{j=1}^m \int_0^\infty f(t,j) dH_j(t)$ 
 is given by 
 \begin{eqnarray*}
   \lefteqn{
     \kappa_{(x,i)}^{(1)}(f) = ( (I-\Gamma )^{-1} f ) (x,i)
   }
   \\
      & = & f(x,i) + \sum_{n=1}^\infty
  \sum_{j_1,\ldots , j_n=1}^m
  \int_0^\infty \cdots \int_0^\infty f(x+y_1+\cdots + y_n,j_n)
  \gamma_{i,j_1} ( dy_1) \cdots \gamma_{j_{n-1},j_n} ( dy_n), 
\end{eqnarray*} 
 $x\geq 0$, $i=1,\ldots , m$.
\item
 For $n\geq 2$, the joint cumulants $\kappa_{(x,i)}^{(n)}(f_1,\ldots , f_n)$ 
 are given by the induction relation 
\begin{align} 
  \label{fjkl}
 & \kappa_{(x,i)}^{(n)}(f_1,\ldots , f_n) 
 = 
 \sum_{k=2}^n
 \sum_{\pi_1\cup \cdots \cup \pi_k = \{1,\ldots , n\}} 
 \left(
 (I-\Gamma )^{-1} \Gamma
 \prod_{j=1}^k \kappa_{(\cdot , \cdot )}^{(|\pi_j|)} ((f_l)_{l\in \pi_j})
 \right)
 (x,i)
,
\end{align} 
$x\geq 0$, $i=1,\ldots , m$,
$n \geq 2$, where the above sum is over set partitions
$(\pi_1,\ldots , \pi_k)$ of $\{1,\ldots , n\}$ 
and $|\pi_i|$ denotes the cardinality of the set $\pi_i$,
 $i=1,\ldots , k$. 
\end{enumerate}
\end{prop}
\noindent
{\em Proof.} See Appendix~\ref{s7}.

\smallskip 
 
\noindent
 Standard (i.e. unconditional) cumulants can then
be obtained in the next corollary as a consequence
of Proposition~\ref{p1}. 
\begin{corollary} 
\label{c1} 
\noindent
The joint cumulants
 $\kappa^{(n)} (f_1,\ldots , f_n)$ 
of $ 
\left(
\sum_{j=1}^m \int_0^\infty f_i(t,j) dH_j(t)
\right)_{1 \leq i \leq n}$
 are given by the relation 
 \begin{align}
   \label{al} 
 & 
  \kappa^{(n)} (f_1,\ldots , f_n)
 = 
 \sum_{i=1}^m
 \sum_{k=1}^n
 \sum_{\pi_1\cup \cdots \cup \pi_k = \{1,\ldots , n\}} 
 \int_0^\infty \prod_{j=1}^k
 \kappa_{(x,i)}^{(|\pi_j|)}((f_i)_{i\in \pi_j})
 \nu_i ( x ) dx, \quad n \geq 1. 
\end{align}
\end{corollary} 
\noindent
{\em Proof.} See Appendix~\ref{s7}.
\subsubsection*{Exponential kernels} 
Joint cumulants will be computed using sums over partitions
and Bell polynomials 
in the case of the exponential offspring intensities 
$$
\gamma_{i,j} (dx) = w_{i,j} {\bf 1}_{[0,\infty )} (x) \re^{-bx} dx, \qquad
  i, j = 1,\ldots , m,
  $$
   given by the
  $m\times m$ connectivity matrix $W = (w_{i,j})_{1\leq i,j \leq m}$, $|w_{i,j}| < b$, 
  and the constant Poisson intensities $\nu_i (dz) = \nu_i dz$, $\nu_i >0$,
  $i,j = 1,\ldots , m$. 
 In this case, the integral operator $\Gamma$ satisfies 
$$
 ( \Gamma f ) (x,i) = \sum_{j=1}^m w_{i,j} \int_0^\infty f(x+y ,j) \re^{-by} dy,
 \quad x\in \real_+, \ i =1,\ldots , n. 
$$ 
 The recursive calculation of joint cumulants
 can be performed using the family of functions
 $e_{p,\eta,t,j}(x,i) := {\bf 1}_{\{ i = j \}} x^p \re^{\eta x} {\bf 1}_{[0,t]}(x)$,
 $\eta < b$, $p\geq 0$, by evaluating $(I-\Gamma )^{-1} \Gamma$
 in Proposition~\ref{p1} on the family of functions $e_{p,\eta,t,j}$ as in the next lemma.
 \begin{lemma}
   \label{l1} 
   For $f$ in the linear span generated by the functions
   $e_{p,\eta, t,k}$, $p\geq 0$, $\eta < b$, $k=1,\ldots , m$, 
   the operator $( I - \Gamma )^{-1} \Gamma$ is given by
$$ 
 ( ( I - \Gamma )^{-1} \Gamma f ) ( x , i) 
 = 
 \sum_{j=1}^m
 \int_0^{t-x}
 f(x+y,j)
 \big[ W \re^{ y W } \big]_{i,j} 
 \re^{-by}
 dy,
 \quad
 x\in [0,t],
 \ 
 i = 1 ,\ldots , m.
$$ 
 \end{lemma}
 For $f$ as in Lemma~\ref{l1}, by Proposition~\ref{p1}
 the first cumulant of
 $\displaystyle \int_0^\infty f_i(t,j) dH_j(t)$
 given that the multidimensional Hawkes process 
 is started from a single jump located in
 $H_i(t)$ at time $x\in \real_+$, $i=1,\ldots , m$,
 is given by 
$$ 
  \kappa_{(x,i)}^{(1)}\big( f( \cdot ) {\bf 1}_{\{j\}} \big)
      = 
 f(x) {\bf 1}_{\{i=j\}}+  \int_0^{t-x}
  \hskip-0.5cm
   \re^{-by}
 f(x+y) \big[ W \re^{ y W } \big]_{i,j} 
 dy, 
$$ 
 $x\in [0,t]$, $i=1,\ldots , m$, and for $n\geq 2$ we have the recursion 
$$ 
 \kappa_{(x,i)}^{(n)}(f{\bf 1}_{[0,t]}) 
  = 
 \sum_{k=2}^n
 \sum_{j=1}^m
 \int_0^{t-x}
 \re^{-by}
 \big[ W \re^{ y W } \big]_{i,j} 
 B_{n,k}
 \big( \kappa_{(x+ y ,j)}^{(1)}(f), \ldots , \kappa_{(x+ y ,j)}^{(n-k+1)} (f)\big) dy. 
$$ 
 The conditional multivariate joint cumulants of
$  
\left(
\int_0^\infty f_i(t,j_i) dH_{j_i}(t)
\right)_{1\leq i \leq n}$
 are given by 
\begin{eqnarray} 
\nonumber 
\lefteqn{
   \kappa_{(x,i)}^{(n)}\big(f_1{\bf 1}_{[0,t_1]} {\bf 1}_{\{j_1\}},\ldots , f_n{\bf 1}_{[0,t_n]} {\bf 1}_{\{j_n\}}\big) 
}
\\
\nonumber 
& = & 
 \sum_{j=1}^m
 \sum_{k=2}^n
 \sum_{\pi_1\cup \cdots \cup \pi_k = \{1,\ldots , n\}} 
  \hskip-0.1cm
 \int_0^{\min(t_1,\ldots , t_m)-x}
  \hskip-0.5cm
   \re^{-by}
 \big[ W \re^{ y W } \big]_{i,j} 
  \prod_{l=1}^k
  \kappa_{(x+y,j)}^{(|\pi_l|)} \big( \big( f_p {\bf 1}_{\{j_p\}}\big)_{p\in \pi_l} \big)
 dy, \qquad 
\end{eqnarray} 
$j_1,\ldots , j_n \geq 1$, with, for $n=2$, 
\begin{align}
\nonumber 
 & \kappa_{(x,i)}^{(2)}\big( f_1 {\bf 1}_{[0,t_1]} {\bf 1}_{\{j_1\}}, f_2 {\bf 1}_{[0,t_2]} {\bf 1}_{\{j_2\}}\big) 
 = 
 \sum_{j=1}^m
 \hskip-0.1cm
 \int_0^{\min ( t_1, t_2 ) -x}
  \hskip-1cm
   \re^{-by}
   \big[ W \re^{ y W } \big]_{i,j} 
 \kappa_{(x+y,j)}^{(1)} \big( f_1 {\bf 1}_{\{j_1\}} \big)
 \kappa_{(x+y,j)}^{(1)} \big( f_2 {\bf 1}_{\{j_2\}} \big)
 dy.  
\end{align} 

\section{Numerical examples}
\label{s3} 
 We consider a nonlinear multivariate Hawkes process
 $\big(\widetilde{H}_1(t),\ldots , \widetilde{H}_m(t)\big)_{t\in \real_+}$ 
 with intensities 
   \begin{equation}
     \label{fjl1} 
   \widetilde{\lambda}_i(t) := \left(
   \nu_i(t) + \sum_{j=1}^n \int_0^t \gamma_{i,j} ( t-s) d \widetilde{H}_j(s)
   \right)^+, \qquad t\in \real_+, 
\end{equation} 
 with exponential offspring intensities 
$$
\gamma_{i,j} (dx) = w_{i,j} {\bf 1}_{[0,\infty )} (x) \re^{-bx} dx, \qquad
  i, j = 1,\ldots , m,
  $$
 where $(w_{i,j})_{1\leq i , j \leq m}$ is a matrix of synaptic weights
 which are possibly negative due to inhibition.
 The inputs  
   $$
   \nu_i(t) + \sum_{j=1}^n \int_0^t \gamma_{i,j} ( t-s) d \widetilde{H}_j(s),
   \qquad
   i=1,\ldots , m, 
   $$
   have been interpreted in \cite{ocker}
   as a family of free neuronal membrane potentials,
   which have the ability to directly influence the
   underlying spike rate.    

   \medskip

   In this paper, we 
   model the membrane potentials $V_i(t )$ 
   using the filtered processes 
$$
V_i (t ) = \int_0^t g_i ( t - s ) dH_i (s), \quad t\in \real_+, 
\quad i =1, \ldots , m, 
$$
where $(H_1(t),\ldots , H_n (t))_{t\geq 0}$ is the multivariate linear Hawkes
process defined in \eqref{fjlkf1}, 
$g_i (t)$ are impulse response functions 
 such that $g_i (u ) = 0$ for $u <0$, $i,j=1,\ldots , m$. 
  We assume that the kernel $g_i (t)$ takes the form 
$$
 g_i (t) := {\bf 1}_{[0,\infty )}(t) \re^{-t/\tau_s}, \qquad i=1,\ldots , m,
   \quad
  t\in \real. 
$$
   We note that although Proposition~\ref{p1} and Corollary~\ref{c1}
   are only proved for $({H}_1(t),\ldots ,$\\ $ {H}_m(t))_{t\in \real_+}$
   with non-negative weights in the cluster process framework of \cite{hawkes},
   the results remain numerically accurate and consistent with the
   sampled cumulants of \eqref{fjl1}, provided that the inhibitory 
   weights $w_{i,j}$ do not become too negative, see \S~1 of \cite{ocker}.
   
   \medskip 
   
   Our cumulant expressions are compared to the sampled cumulants of the
   nonlinear Hawkes process 
   $\big(\widetilde{H}_1(t),\ldots , \widetilde{H}_m(t))_{t\in \real_+}$
   in \eqref{fjl1} in the presence of negative weights.
 The joint cumulant  
 $\langle \langle V_{l_1}(t_1)\cdots V_{l_n}(t_n) \rangle \rangle$, 
 $1 \leq l_1,\ldots , l_n \leq m$, is evaluated in closed form by induction
 by the command 
 $\verb|c(W,b,[g,...,g],[l1,...,ln],[t1,...,tn])|$ 
 in Maple, or 
 $\verb|c[W,b,{g,...,g},{l1,...,ln},{t1,...,tn}]|$
 in Mathematica, 
 defined in the code blocks presented in Appendix~\ref{a2}. 
 Closed form expressions for higher order joint moments and cumulants 
 may involve thousands of terms resulting of symbolic computations in Maple
 or Mathematica, nevertheless their numerical
 implementation remains attractive in terms of computation time and
 stability properties. 
 
 \medskip

 In the following numerical examples we take $m=4$ and consider
 the potentials
 $(V_1(t),V_2(t),V_3(t),V_4(t))= (V_{\scaleto{\rm E1}{4.4pt}} (t),V_{\scaleto{\rm E2}{4.4pt}} (t),V_{\scaleto{\rm E3}{4.4pt}} (t),V_{\scaleto{\rm I}{4.4pt}} (t))$ 
 with three excitatory neurons and one inhibitory neuron,
 parametrized by the weight matrix
$$ 
W = (w_{i,j})_{1\leq i,j \leq 4}
=
\bbordermatrix{
    ~& {\rm E1} ~& {\rm E2} ~& {\rm E3} ~& {\rm I} \cr
 {\rm E1}  ~& 10 ~& 0 ~& 10 ~& 0 \cr
 {\rm E2}  ~& 0 ~& 10 ~& 10 ~& -8 \cr
 {\rm E3}  ~& 10 ~& 10 ~& 0 ~& -8 \cr
 {\rm I}  ~& 10 ~& 10 ~& 10 ~& -10 \cr
}. 
$$ 
 Note that this example does not have reset-like effects. 
Although this example is restricted to four neurons
for the sake of computation time, the algorithm is valid for any
 $m\geq 1$. The connectivity of the network can be represented as follows. 
 
\vspace{-0.4cm}

\begin{equation} 
\nonumber 
\begin{tikzpicture}[->,>=stealth',shorten >=1pt,auto,node distance=2cm,scale=0.75,semithick]
\tikzstyle{VertexStyle}=[shape = circle, fill = blue!20, minimum size = 28pt, text = black, draw]
\Vertex[x=-2, y=3]{E1}
\Vertex[x=9, y=3]{E2}
\Vertex[x=4, y=0]{E3}
\Vertex[x=4, y=6]{I}
\path[color=black]   (E1) edge  [loop left] node {10} (E1);
\path[color=black]   (E1) edge  [bend right] node[below] {10} (E3);
\path[color=black]   (E2) edge  [bend right] node[above] {-8} (I);
\path[color=black]   (E2) edge  [loop right] node {$10$} (E2);
\path[color=black]   (E2) edge  [bend right] node[below] {10} (E3);
\path[color=black]   (E3) edge  [bend right] node[below] {10} (E1);
\path[color=black]   (E3) edge  [bend right] node[above] {10} (E2);
\path[color=black]   (E3) edge  [bend right] node[left] {-8} (I);
\path[color=black]   (I) edge  [bend right] node[below] {10} (E1);
\path[color=black]   (I) edge  [bend right] node[above] {10} (E2);
\path[color=black]   (I) edge  [bend right] node[left] {10} (E3);
\path[color=black]   (I) edge  [loop above] node[left] {-10} (I);
\end{tikzpicture}
\end{equation}
 
\begin{figure}[H]
\centering
\hskip-0.1cm
\begin{subfigure}{.49\textwidth}
\centering
\includegraphics[width=1\textwidth]{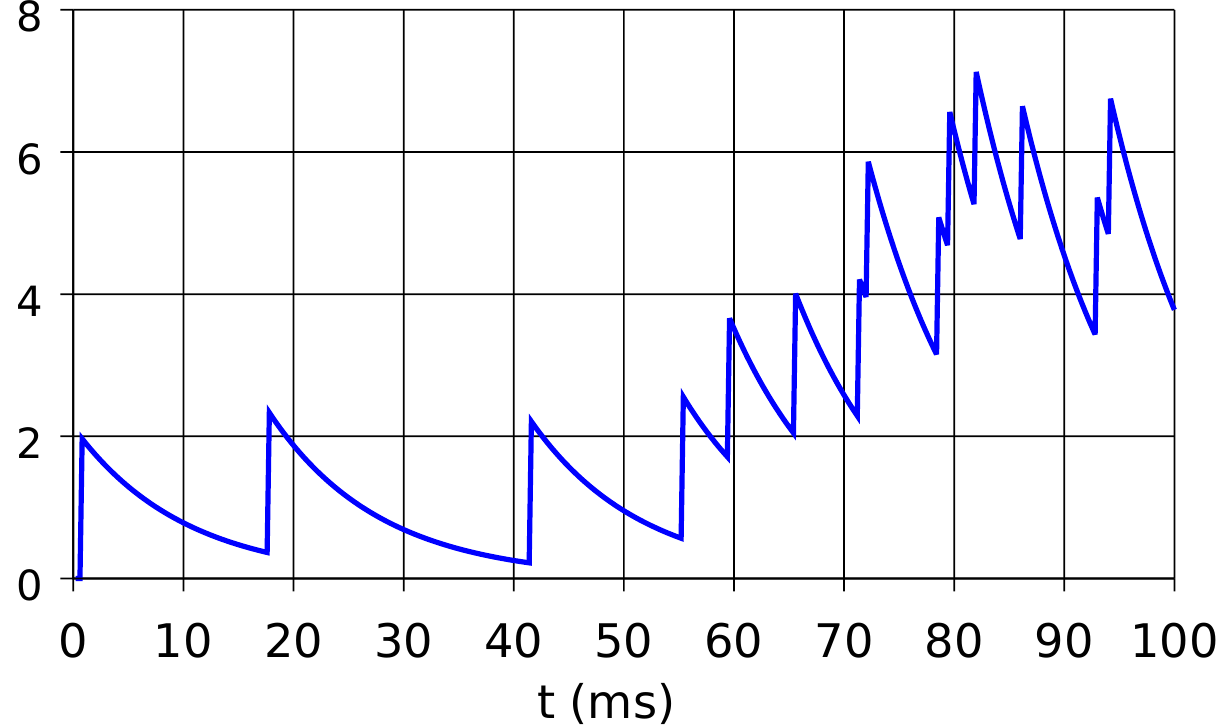} 
\caption{\small Excitatory potential $V_2(t)=V_{\scaleto{\rm E2}{4.4pt}} (t)$.} 
\end{subfigure}
\begin{subfigure}{.49\textwidth}
\centering
\includegraphics[width=1\textwidth]{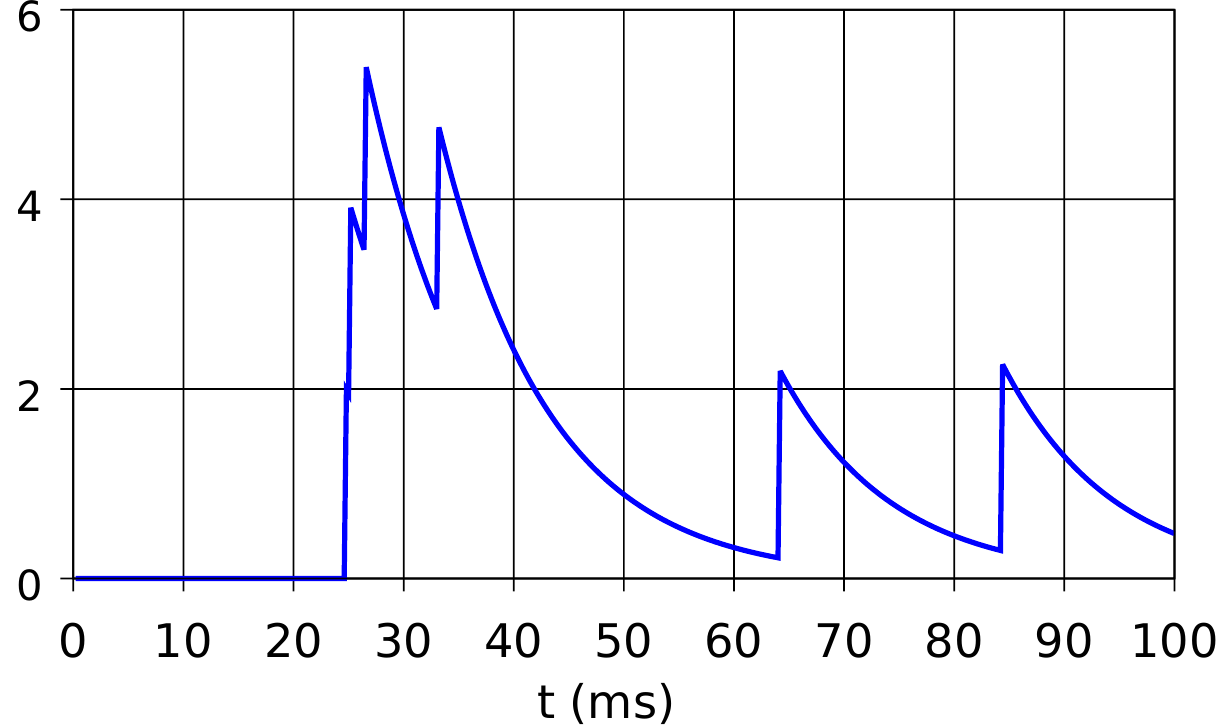} 
\caption{\small Inhibitory potential $V_4(t)=V_{\scaleto{\rm I}{4.4pt}} (t)$.} 
\end{subfigure}
\caption{\small Filtered shot noise processes.} 
\label{fig4}
\end{figure}

\noindent
 Figure~\ref{fig4} presents random simulations
 of the membrane potentials $V_{\scaleto{\rm E2}{4.4pt}} (t )$ and $V_{\scaleto{\rm E4}{4.4pt}} (t)$ 
 with 
 $T=0.1$ seconds with $g(u) = \re^{-u/\tau_s } {\bf 1}_{[0,\infty )}(u)$,
   with 
   $b=50$Hz,
   $\tau_s = 0.01$ seconds and 
   $\nu_i = 250$Hz, $i=1,2,3,4$. 
   We use the algorithm of \cite{ogata2} for the
   simulation of multivariate Hawkes processes,
   and its implementation given in \cite{ychen2}. 

\vspace{-0.05cm}
 
\vspace{-0.3cm}

\noindent
The following Figures~\ref{fig4.11-0} to \ref{fig4.22-0}
presents numerical cumulant estimates using closed form
expression, and compares them with Monte Carlo simulations
run with 10 million samples.
Figure~\ref{fig4.11-0} presents numerical estimates of first moment
and standard deviation,  
together with the mean obtained by Monte Carlo simulations.

\vskip-0.1cm

\begin{figure}[H]
\centering
\hskip-0.1cm
\begin{subfigure}{.49\textwidth}
\centering
\includegraphics[width=1.\textwidth]{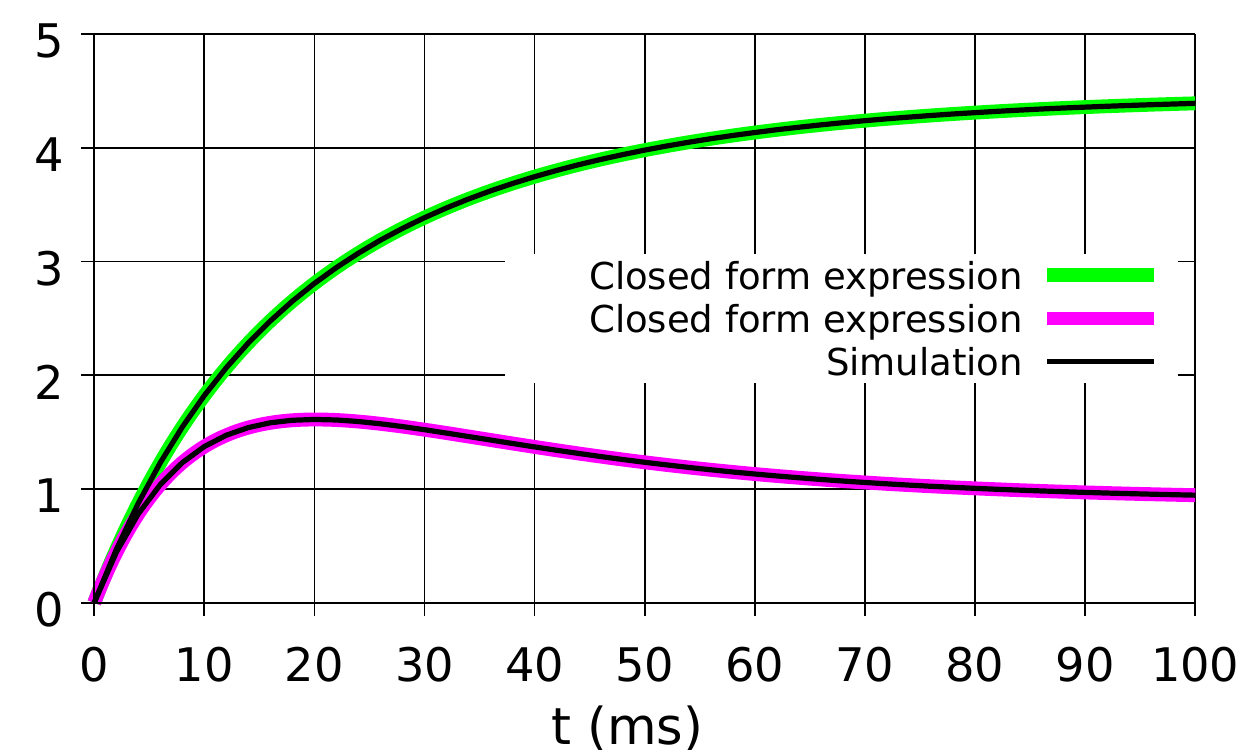} 
\caption{\small Means of $V_2(t)=V_{\scaleto{\rm E2}{4.4pt}} (t)$ and $V_4(t)=V_{\scaleto{\rm I}{4.4pt}} (t)$.} 
\end{subfigure}
\hskip0.1cm
\begin{subfigure}{.49\textwidth}
\centering
\includegraphics[width=1.\textwidth]{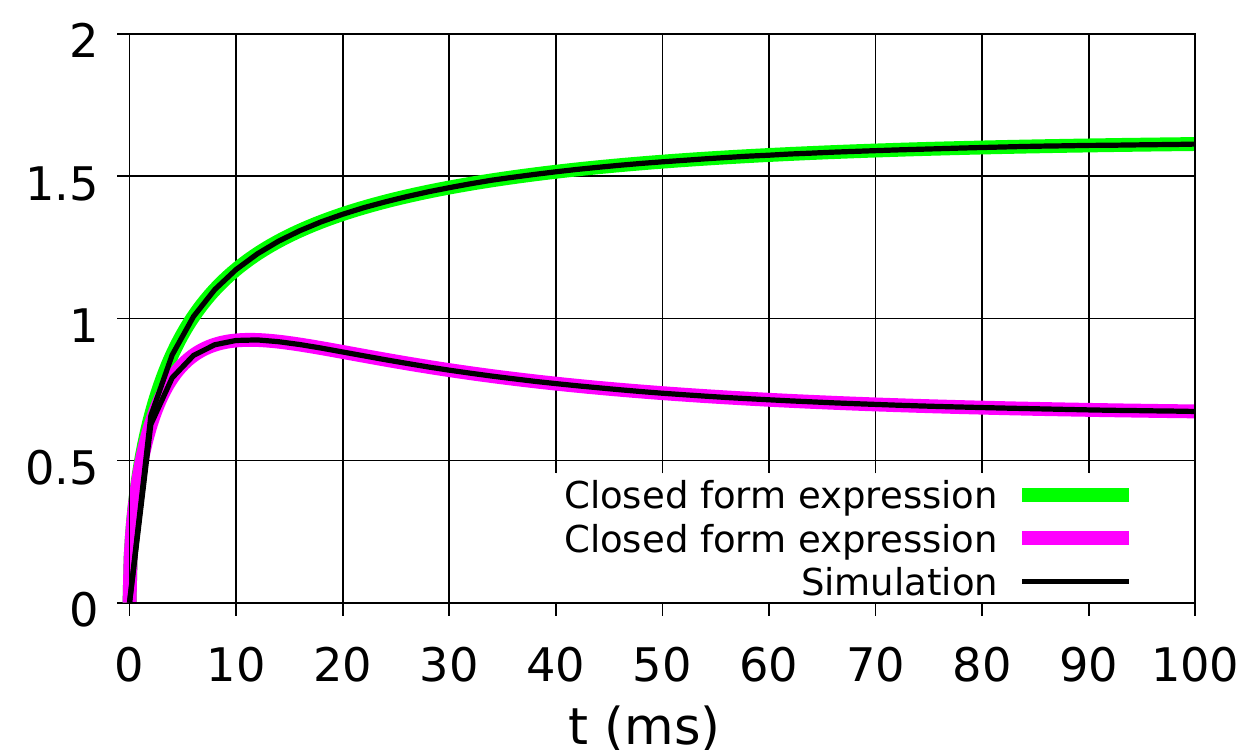} 
\caption{\small Standard deviations of $V_{\scaleto{\rm E2}{4.4pt}} (t)$ and $V_{\scaleto{\rm I}{4.4pt}} (t)$.} 
\end{subfigure}
\caption{\small Excitatory and inhibitory means and standard deviations.} 
\label{fig4.11-0}
\end{figure}

\vspace{-0.3cm}

\noindent
Figures~\ref{fig4.11-0}-\ref{fig4.11}
can be obtained from the Maple commands listed below together with their runtimes
on a standard laptop computer, 
after loading the function definitions listed in Appendix~\ref{a2}
and the variable assignments of $W$ and $\mu$.

\begin{table}[H] 
  \centering
\scriptsize 
  \begin{tabular}{|l|l|c|}
 \hline
 \multicolumn{3}{|l|}{
   W := $<<10, 0, 10, 10> | <0, 10, 10, 10> | <10, 10, 0, 10> | <0, -8, -8, -10>>$;
   }
\\
 \hline
 \multicolumn{3}{|l|}{
mu := [t -> 250, t -> 250, t -> 250, t -> 250]; g := (x, t) -> exp(-100*t + 100*x);
} 
\\
\hline
\hline
Instruction & Computed quantity & Computation time
 \\ 
 \hline
 \hline
 c(W, 50, [g], [2], [t], mu) & First cumulant of V2(t) & One second 
\\
\hline
c(W, 50, [g,g], [4,4], [t,t], mu) & Second cumulant of V4(t) &  7 seconds 
\\
\hline
c(W, 50, [g,g], [4,2], [t,0.05], mu) &  Covariance of (V2(t1),V4(t)) for t<t1=0.05 & 12 seconds 
\\
\hline
c(W, 50, [g,g], [2,4], [0.05,t], mu) &  Covariance of (V2(t1),V4(t)) for t>t1=0.05 &  15 seconds 
 \\ 
\hline
\end{tabular}
\end{table} 

\vskip-0.3cm

\noindent 
 Figure~\ref{fig4.11} presents estimates of
 the cross-correlations ${\rm Cor} ( V_{\scaleto{\rm E2}{4.4pt}} (t), V_{\scaleto{\rm E4}{4.4pt}} (t))$
 and ${\rm Cor} ( V_{\scaleto{\rm E2}{4.4pt}} (t_1), V_{\scaleto{\rm I}{4.4pt}} (t))$ 
 with $t_1 := 50$ms and $t\in [0,10{\rm ms}]$. 

\vskip-0.1cm

\begin{figure}[H]
\centering
\hskip-0.1cm
\begin{subfigure}{.49\textwidth}
\centering
\includegraphics[width=1.\textwidth]{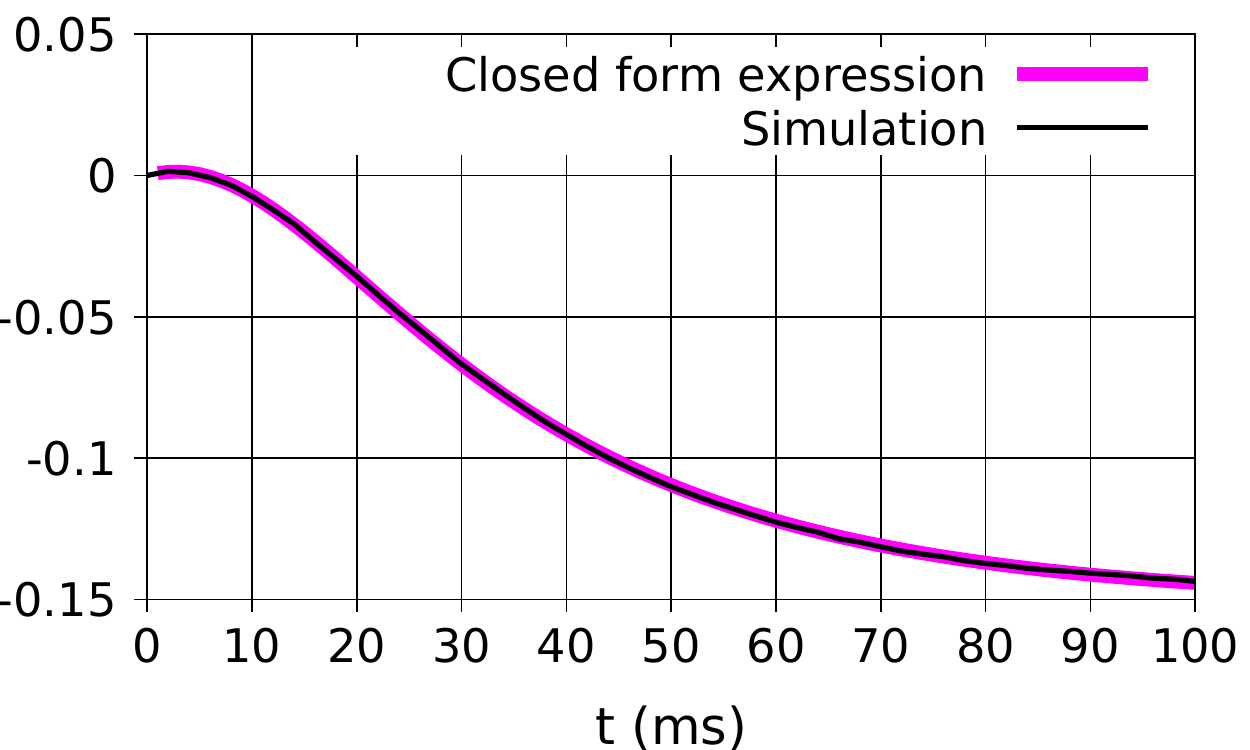} 
\caption{\small Cross-correlation of $(V_{\scaleto{\rm E2}{4.4pt}} (t),V_{\scaleto{\rm I}{4.4pt}} (t))$.}
\end{subfigure}
\hskip0.1cm
\begin{subfigure}{.49\textwidth}
\centering
\includegraphics[width=1.\textwidth]{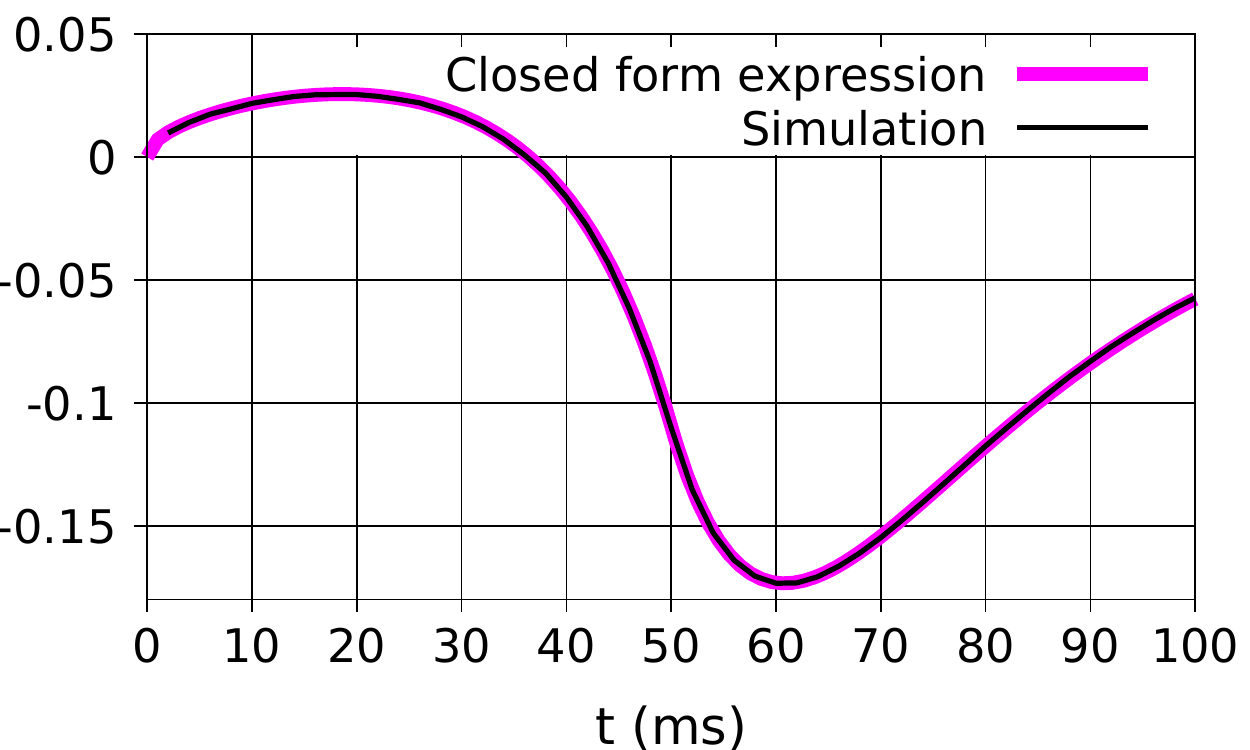} 
\caption{\small Cross-correlation of $(V_{\scaleto{\rm E2}{4.4pt}} (t_1),V_{\scaleto{\rm I}{4.4pt}} (t))$.} 
\end{subfigure}
\caption{\small Cross-correlations of $(V_{\scaleto{\rm E2}{4.4pt}} (t),V_{\scaleto{\rm I}{4.4pt}} (t))$ and $(V_{\scaleto{\rm E2}{4.4pt}} (t_1),V_{\scaleto{\rm I}{4.4pt}} (t))$ with $t_1=50ms$.} 
\label{fig4.11}
\end{figure}

\vspace{-0.3cm}

\noindent
Figure~\ref{fig400-3} presents time-dependent estimates of the third cumulant
of $V_{\scaleto{\rm I}{4.4pt}}(t)$ and third joint cumulant of $(V_{\scaleto{\rm E1}{4.4pt}}(t_1),V_{\scaleto{\rm E1}{4.4pt}}(t_1),V_{\scaleto{\rm I}{4.4pt}} (t))$ 
 with $t_1 = 0.05$, based on the exact moment expressions 
 computed in Maple by the following commands. 

\begin{table}[H] 
  \centering
\scriptsize 
  \begin{tabular}{|l|l|c|}
 \hline
Instruction & Computed quantity & Computation time
 \\ 
 \hline
 \hline
c(W, 50, [g,g,g], [4,4,4], [t,t,t], mu) & Third cumulant of V4(t) & 56 seconds 
 \\ 
 \hline
c(W, 50, [g,g,g], [4,1,1], [t,0.05,0.05], mu) & Third joint cumulant of (V1(t1),V1(t1),V4(t)), t<0.05  & 239 seconds
 \\ 
 \hline
c(W, 50, [g,g,g], [1,1,4], [0.05,0.05,t], mu) & Third joint cumulant of (V1(t1),V1(t1),V4(t)), t>0.05 & 473 seconds
 \\ 
\hline
\end{tabular}
\end{table} 

\vspace{-0.7cm}

\begin{figure}[H]
\centering
\hskip-0.25cm
\begin{subfigure}{.49\textwidth}
\centering
\includegraphics[width=1.0\textwidth]{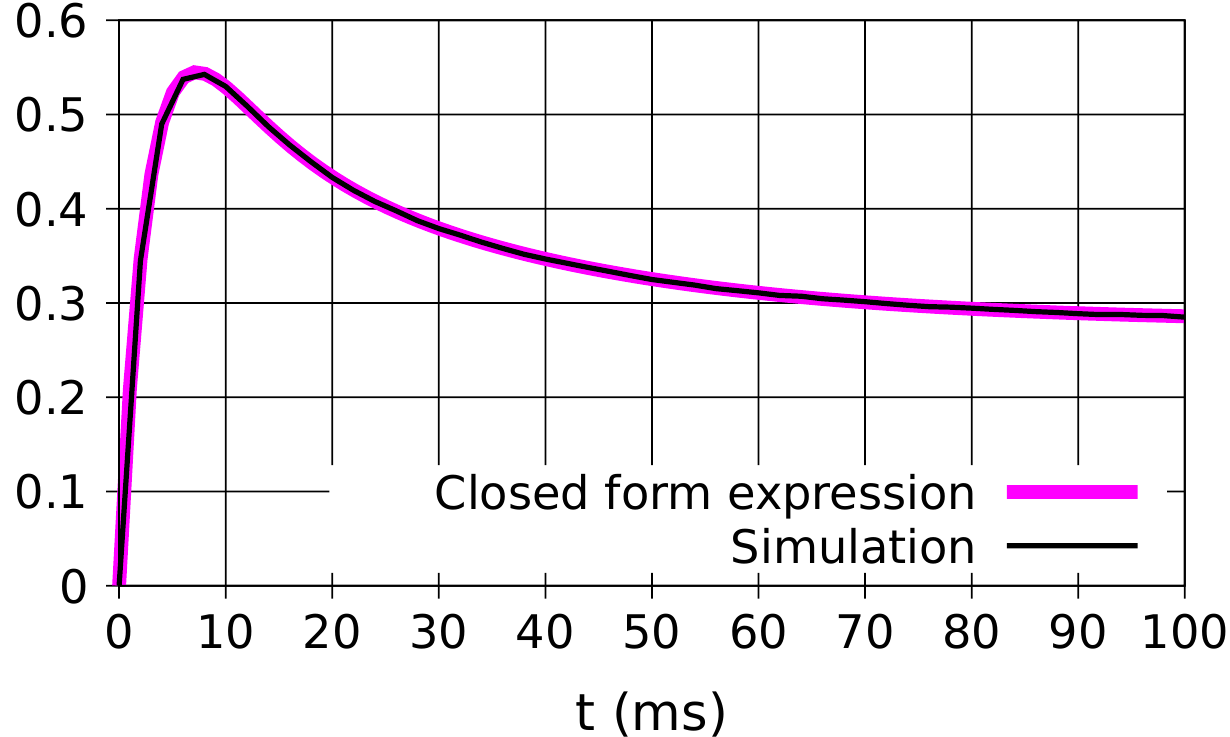}
\caption{\small Third cumulant of $V_4(t)=V_{\scaleto{\rm I}{4.4pt}}(t )$.} 
\end{subfigure}
\hskip0.cm
\begin{subfigure}{.49\textwidth}
\centering
\includegraphics[width=1.0\textwidth]{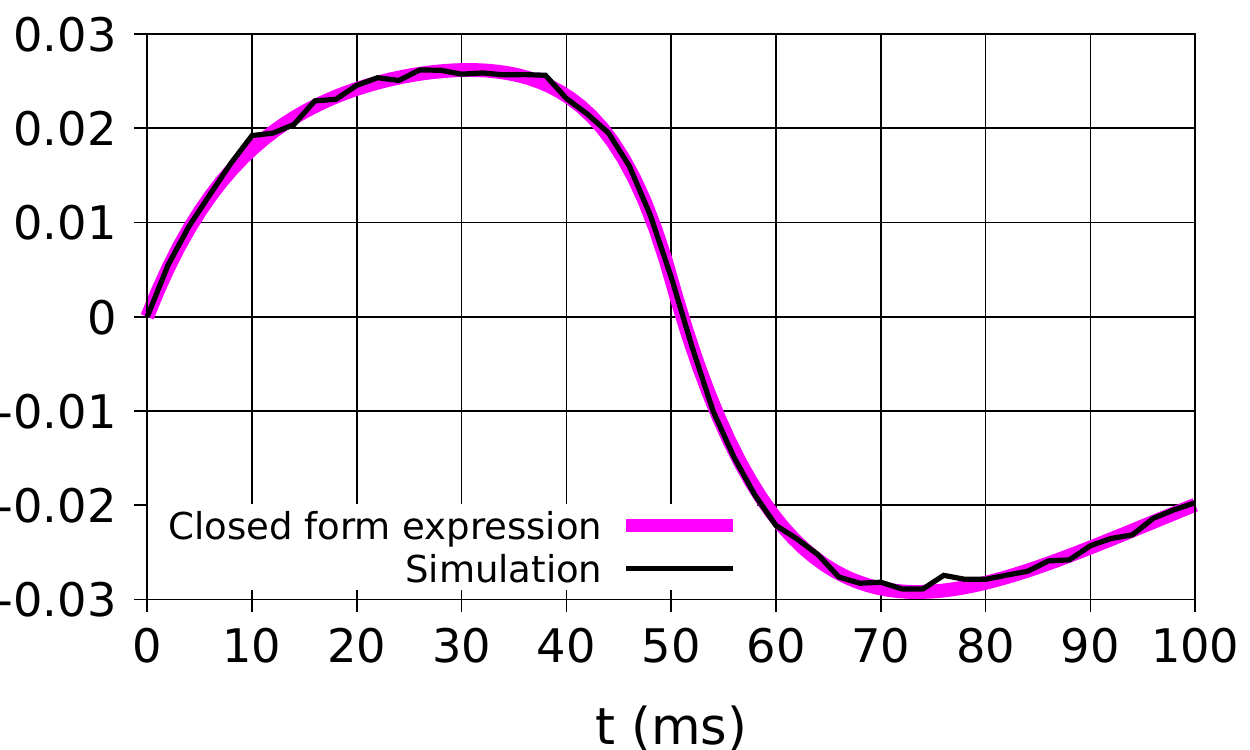}
\caption{\small Joint cumulant of $(V_{\scaleto{\rm E1}{4.4pt}}(t_1),V_{\scaleto{\rm E1}{4.4pt}}(t_1),V_{\scaleto{\rm I}{4.4pt}}(t))$.} 
\end{subfigure}
\caption{\small Third order cumulants with $t_1=50ms$.} 
\label{fig400-3}
\end{figure}

\vspace{-0.3cm}

\noindent
 Figure~\ref{fig4.22-0} presents estimates of 
 the fourth cumulant of $V_{\scaleto{\rm I}{4.4pt}} (t)$ and of the fourth joint cumulant of   
 and $(V_{\scaleto{\rm E1}{4.4pt}} (t ), V_{\scaleto{\rm E2}{4.4pt}} (t ), V_{\scaleto{\rm E3}{4.4pt}} (t ), V_{\scaleto{\rm I}{4.4pt}} (t))$ 
 respectively, computed by the following commands
 
\begin{table}[H] 
\centering
\scriptsize 
\begin{tabular}{|l|l|c|}
 \hline
Instruction & Computed quantity & Computation time
 \\ 
 \hline
 \hline
c(W, 50, [g,g,g,g], [4,4,4,4], [t,t,t,t], mu) & Fourth cumulant of V4(t) & 677 seconds 
 \\ 
 \hline
c(W, 50, [g,g,g,g], [1,2,3,4], [t,t,t,t], mu) & Fourth joint cumulant of (V1(t),V2(t),V3(t),V4(t)) & 14917 seconds
 \\ 
\hline
\end{tabular}
\end{table} 

\vspace{-0.7cm}

\begin{figure}[H]
\centering
\hskip-0.2cm
\begin{subfigure}{.48\textwidth}
\centering
\includegraphics[width=1.\textwidth]{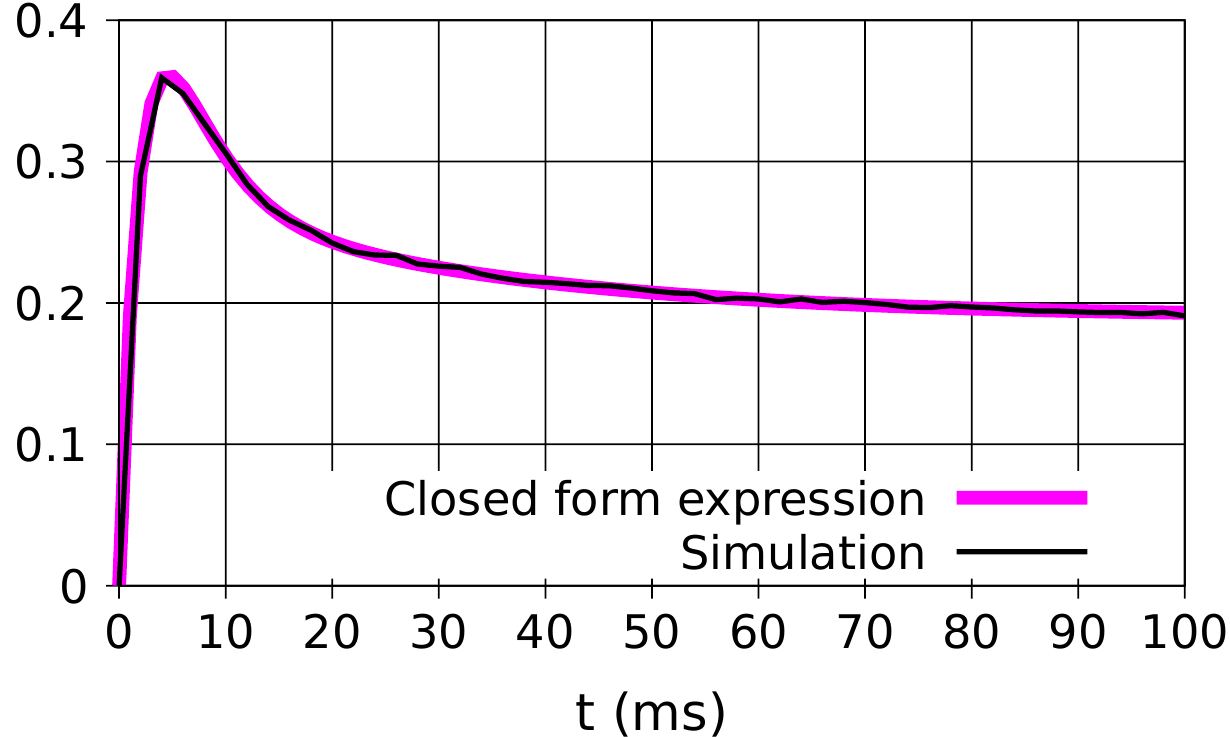} 
\caption{\small Fourth cumulant of $V_4(t)=V_{\scaleto{\rm I}{4.4pt}} (t)$.}
\end{subfigure}
\hskip0.6cm
\begin{subfigure}{.48\textwidth}
\centering
\vskip0.05cm
\includegraphics[width=1.\textwidth]{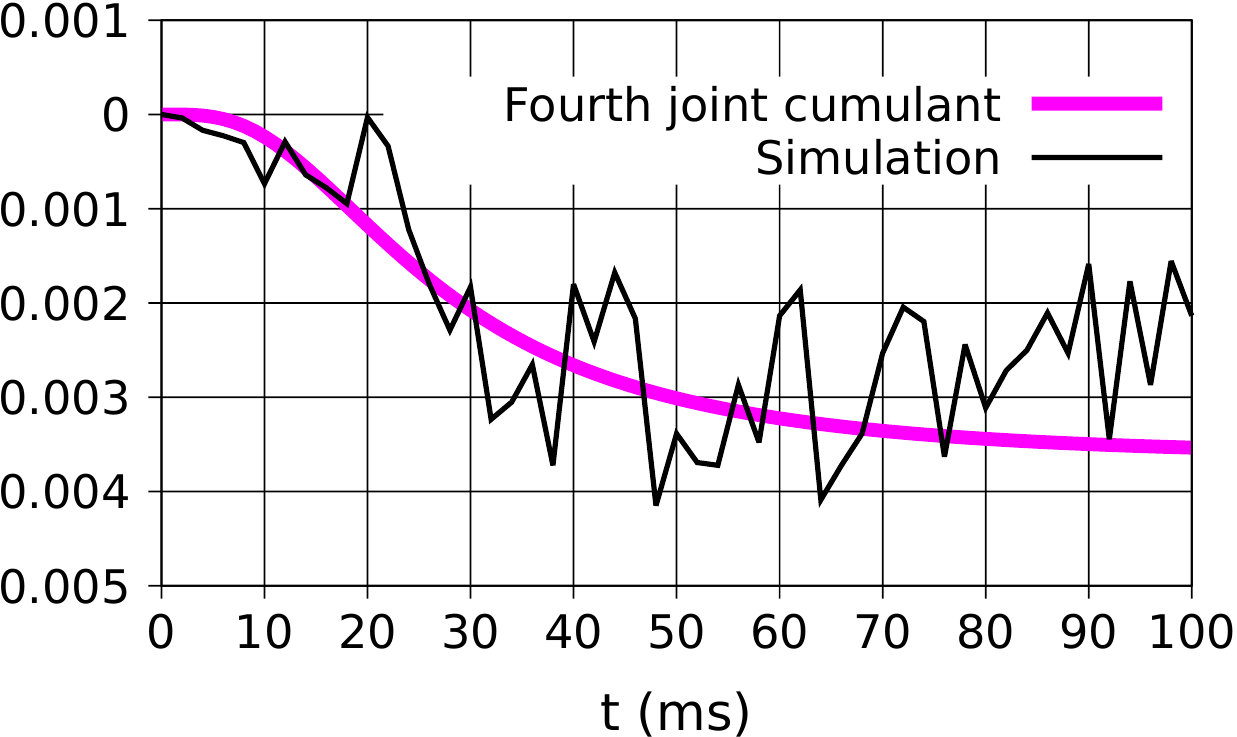} 
\caption{\small Joint cumulant of $(V_{\scaleto{\rm E1}{4.4pt}} (t ),\ldots , V_{\scaleto{\rm E3}{4.4pt}} (t ), V_{\scaleto{\rm I}{4.4pt}} (t ))$.} 
\end{subfigure}
\caption{\small Fourth order cumulants.} 
\label{fig4.22-0}
\end{figure}

\vspace{-0.3cm}

\noindent
One can check from
Figures~\ref{fig400-3}-$b)$ and \ref{fig4.22-0}-$b)$
that the precision of Monte Carlo
    estimation is degraded starting with
    joint third cumulants and fourth cumulants,
    while it becomes clearly insufficient 
    for an accurate estimation of fourth joint cumulants 
    in Figure~\ref{fig4.22-0}-$b)$.
        This phenomenon has also been observed in 
 \cite{neuron} when modeling neuronal activity using Poisson processes, and 
 can be attributed to the fact that the estimation of fourth-order joint cumulants 
 in terms of sampled moments involves a multinomial expression of order
 four in $4$ variables with changing signs. 

    \medskip
    
\begin{figure}[H]
\centering
\hskip-0.2cm
\begin{subfigure}{.48\textwidth}
\centering
\includegraphics[width=1.\textwidth]{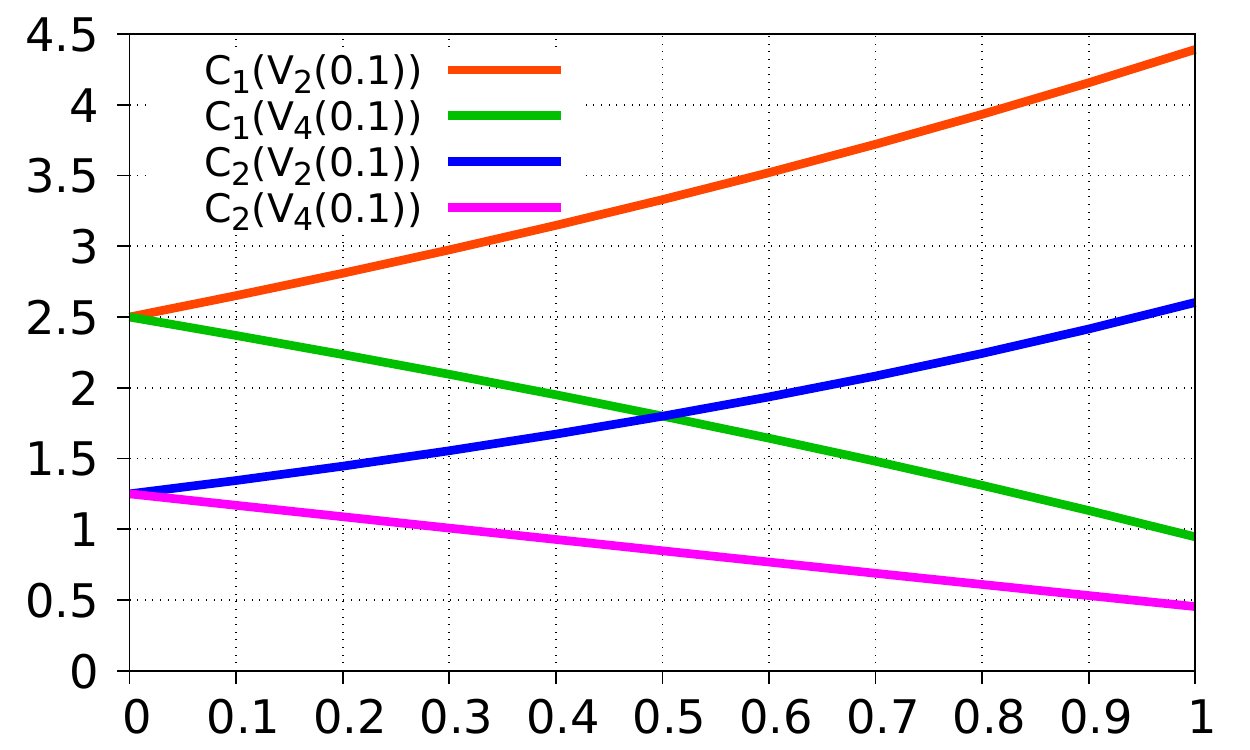} 
\caption{\small Sensitivities of first and second cumulants.} 
\end{subfigure}
\hskip0.6cm
\begin{subfigure}{.48\textwidth}
\centering
\vskip0.05cm
\includegraphics[width=1.\textwidth]{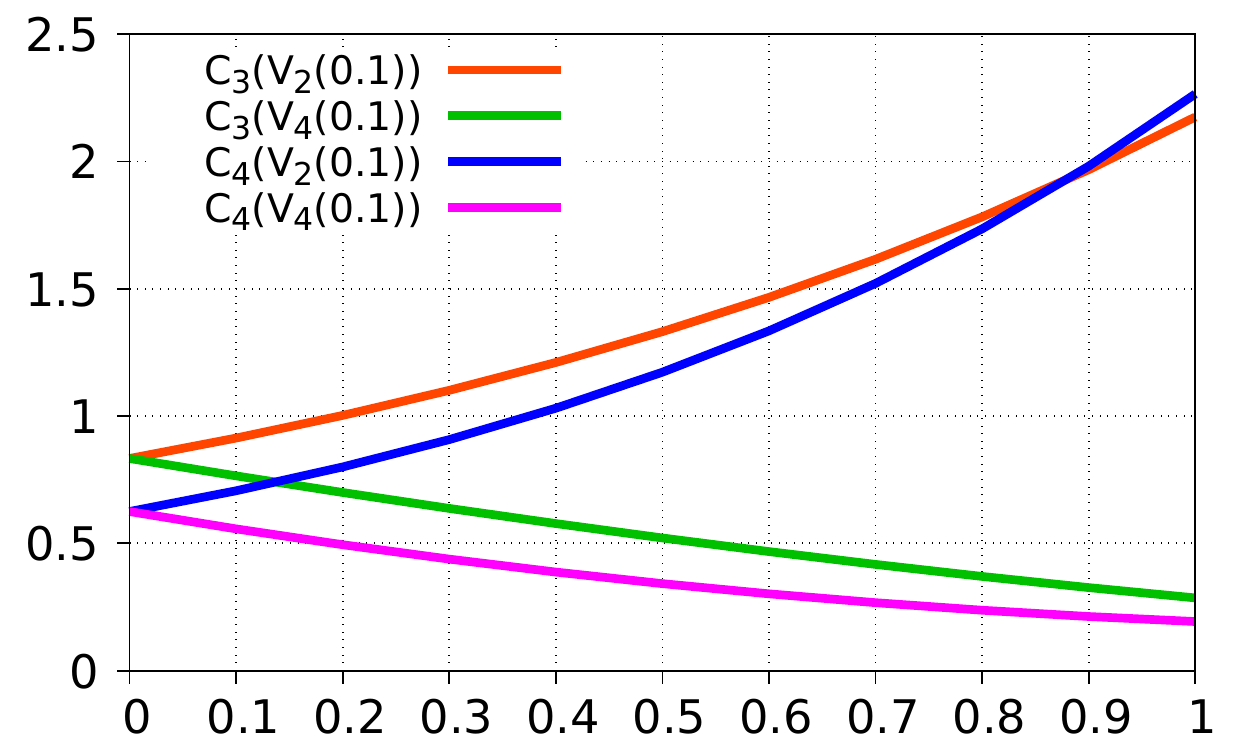} 
\caption{\small Sensitivities of third and fourth cumulants.} 
\end{subfigure}
\caption{\small Sensitivities of cumulants.} 
\label{fig4.22}
\end{figure}

The knowledge of cumulants in explicit form also allows us to study their behavior
under the variation of other parameters. In Figure~\ref{fig4.22} we plot the
respective evolutions of the first four cumulants of $V_{\scaleto{\rm E2}{4.4pt}} (0.1)$ and $V_{\scaleto{\rm I}{4.4pt}} (0.1)$
as a function of $\alpha W$ with $\alpha \in [0,1]$. 

\vspace{-0.3cm}

\section{Gram-Charlier expansions}
\label{s5}
\begin{figure}[H]
\centering
\includegraphics[width=0.785\textwidth,height=5.5cm]{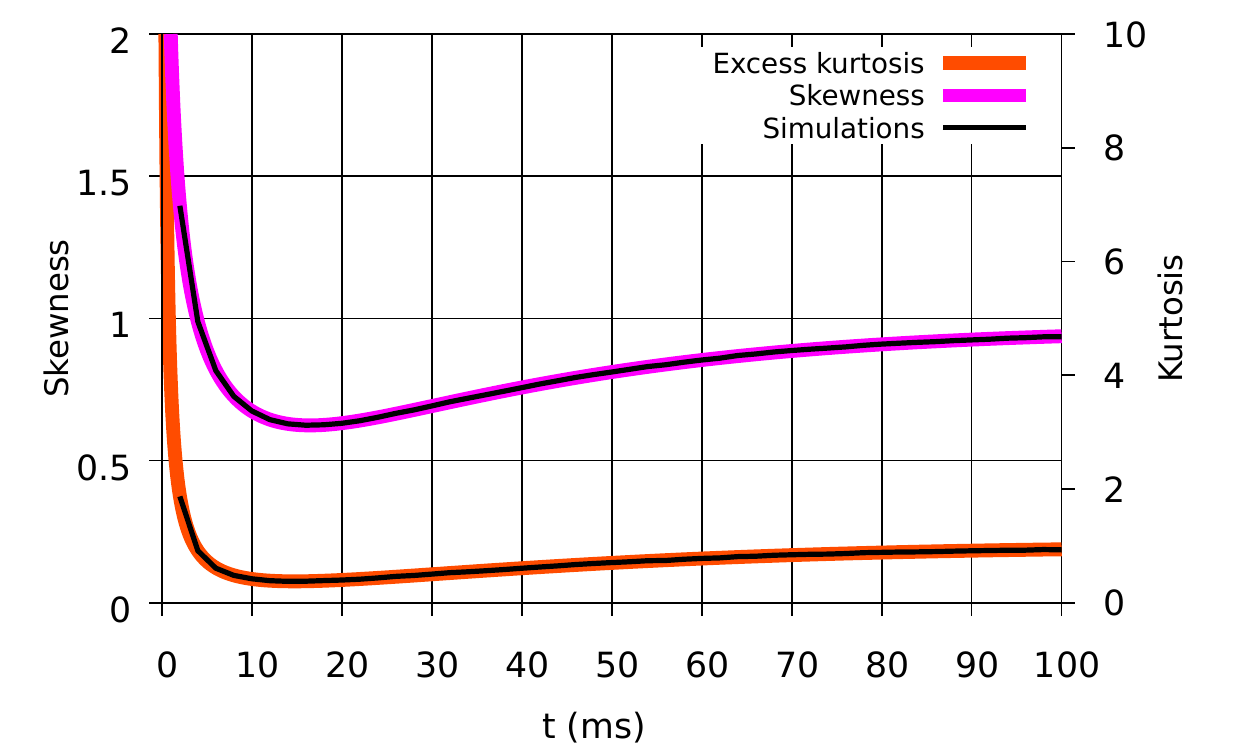} 
\caption{\small Skewness and kurtosis of $V_4(t)=V_{\scaleto{\rm I}{4.4pt}} (t )$.} 
\label{fig4.1b.0}
\end{figure}

In this section we use our cumulant formulas for
the estimation of probability densities of potentials
by Gram-Charlier expansions. 
The Gram-Charlier expansion of the continuous
probability density function
$\phi_X(x)$ of a random variable $X$ is given by 
\begin{equation} 
\label{gram_charlier}
\phi_X(x)=
\frac{1}{\sqrt{\kappa_2}}
\varphi \left( \frac{x-\kappa_1}{\sqrt{\kappa_2}}\right)
+
\frac{1}{\sqrt{\kappa_2}}
\sum_{n=3}^{\infty}
c_n H_n\left(
\frac{x-\kappa_1}{\sqrt{\kappa_2}} \right)
\varphi \left( \frac{x-\kappa_1}{\sqrt{\kappa_2}}\right),
\end{equation}
 see \S~17.6 of \cite{cramer}, where 
\begin{itemize}
  \item $\displaystyle \varphi(x) :=\frac{1}{\sqrt{2\pi}}\re^{-x^2/2}$,
 $ x\in \real$,
    is the standard normal density function,
    \item 
$\displaystyle 
H_n(x):=\frac{(-1)^n}{\varphi(x)}
\frac{\partial^n \varphi}{\partial x^n}(x)$,
 $x\in\real$, 
 is the Hermite polynomial of degree $n\geq 0$, with
$H_0(x)=1$, 
$H_1(x)=x$, 
$H_3(x)=x^3-3x$,
$H_4(x)=x^4-6x^2+3$, 
 $H_6(x)=x^6-15x^4+45x^2-15$,
 \item 
  the sequence $(c_n)_{n\geq 3}$ is given from the cumulants $(\kappa_n)_{n\ge 1}$
of $X$ as 
$$ 
c_n = \frac{1}{(\kappa_2)^{n/2}}
\sum_{m=1}^{[n/3]}
\sum_{\substack{l_1+\cdots+l_m=n\\
{l_1,\ldots , l_m \ge 3}}}\frac{\kappa_{l_1}\cdots \kappa_{l_m}}{m! l_1!\cdots l_m!}, \qquad n\ge 3.
$$ 
\end{itemize} 
 In particular, $c_3$ and $c_4$ can be expressed from 
 the skewness $\kappa_3/(\kappa_2)^{3/2}$ and
 the excess kurtosis $\kappa_4/(\kappa_2)^2$, with 
$$ 
c_3 = \frac{\kappa_3}{3! (\kappa_2)^{3/2}}, 
\quad
c_4 = \frac{\kappa_4}{4! (\kappa_2)^2}, 
\quad 
c_5 = \frac{\kappa_5}{5! \kappa_5^{5/2}},
\quad
\mbox{and}
\quad
 c_6 =
 \frac{\kappa_6}{6! (\kappa_2)^3}
 +
  \frac{(\kappa_3)^2}{2(3!)^2 (\kappa_2)^3}
.
$$
 Figure~\ref{fig4.1b.0} presents numerical estimates of skewness 
 and excess kurtosis of $V_{\scaleto{\rm I}{4.4pt}} (t )$
 obtained from exact cumulant expressions. 
 
 \vskip-0.2cm

\vspace{-0.3cm}

\begin{figure}[H]
\centering
\begin{subfigure}{.5\textwidth}
\centering
\includegraphics[width=1.\textwidth]{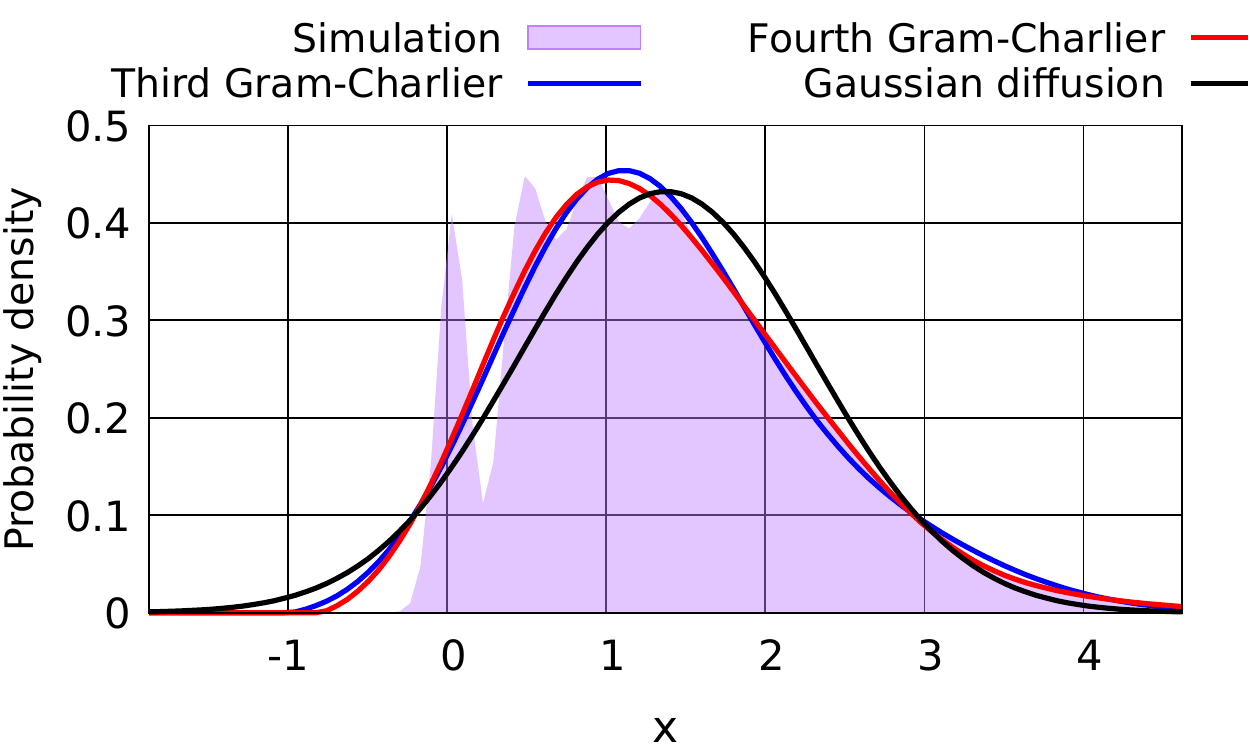} 
\caption{\small t=10 ms.} 
\end{subfigure}
\hskip-0.2cm
\begin{subfigure}{.5\textwidth}
\centering
\includegraphics[width=1.\textwidth]{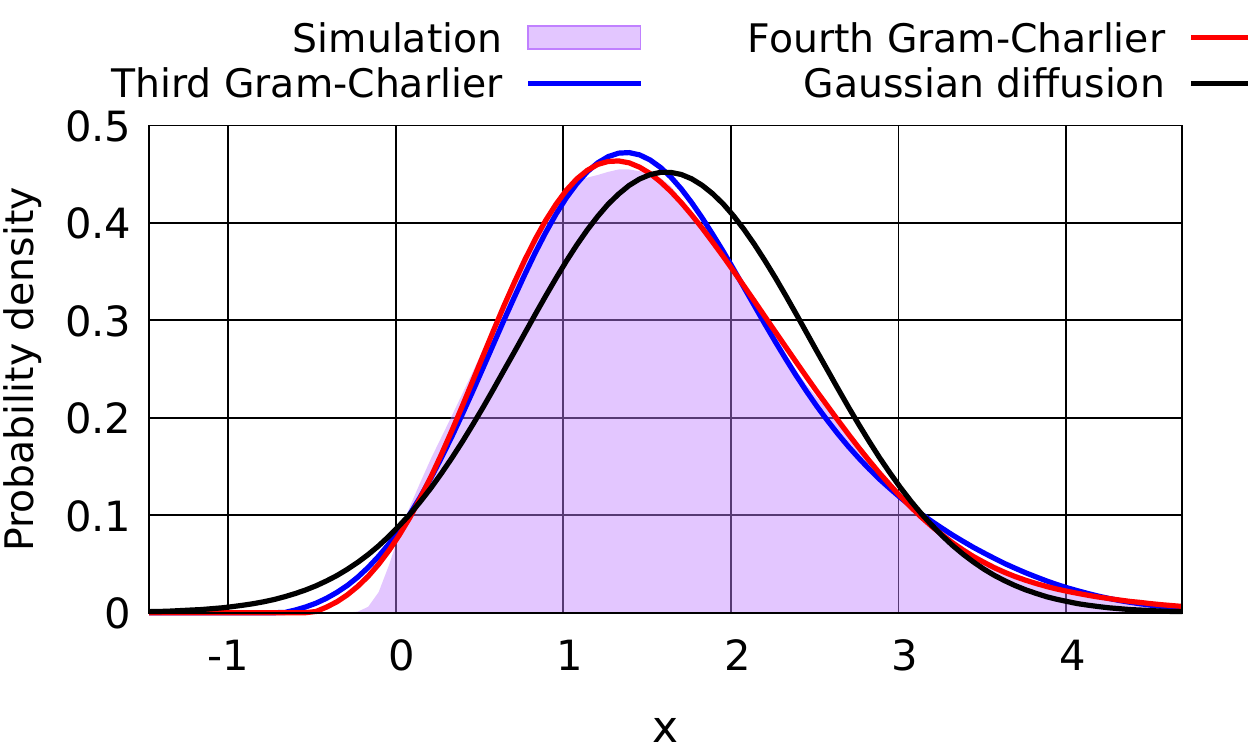} 
\caption{\small t=20 ms.} 
\end{subfigure}
\caption{\small Gram-Charlier density expansions {\em vs} Monte Carlo density estimation.} 
\label{fig5}
\end{figure}

\noindent
 As above, our results, 
 which are only proved for non-negative weights, remain accurate 
 although the considered Hawkes process allows for inhibition. 
 In what follows, we use third and fourth-order expansions given by 
\begin{equation} 
\nonumber 
\phi_X^{(3)}(x)=
\frac{1}{\sqrt{\kappa_2}}
\varphi \left( \frac{x-\kappa_1}{\sqrt{\kappa_2}}\right)
\left( 1 +
c_3 H_3\left(
\frac{x-\kappa_1}{\sqrt{\kappa_2}} \right)
\right)
\end{equation}
 and
\begin{equation} 
\nonumber 
\phi_X^{(4)}(x)=
\frac{1}{\sqrt{\kappa_2}}
\varphi \left( \frac{x-\kappa_1}{\sqrt{\kappa_2}}\right)
\left( 1 +
c_3 H_3\left(
\frac{x-\kappa_1}{\sqrt{\kappa_2}} \right)
+
c_4 H_4\left(
\frac{x-\kappa_1}{\sqrt{\kappa_2}} \right)
 + c_6 H_6\left( \frac{x-\kappa_1}{\sqrt{\kappa_2}} \right)
 \right),
\end{equation}
 and compare them to the first-order expansion
\begin{equation} 
\nonumber 
\phi_X^{(1)}(x)=
\frac{1}{\sqrt{\kappa_2}}
\varphi \left( \frac{x-\kappa_1}{\sqrt{\kappa_2}}\right)
\end{equation}
 which corresponds to a Gaussian diffusion approximation. 
Figure~\ref{fig5} presents 
second, third and fourth-order Gram-Charlier expansions \eqref{gram_charlier}
 for the probability density function of the
membrane potential $V_{\scaleto{\rm I}{4.4pt}} (t )$, based on the exact cumulant expressions  
 computed at the times $t=10$ms and $t=20$ms. 
The purple areas correspond to probability density estimates
obtained by Monte Carlo simulations. 
The second-order expansions correspond to
the Gaussian diffusion approximation 
obtained by matching first and second-order moments. 
  
\vspace{-0.3cm}

\noindent
Figures~\ref{fig4.1b.0} and \ref{fig5} show that
the actual probability density estimates obtained by simulation 
 are significantly different from
their Gaussian diffusion approximations when 
skewness and kurtosis take large absolute values. 
In addition, in Figure~\ref{fig5} 
the fourth-order Gram-Charlier expansions appear to give the best fit
to the actual probability densities, 
which have negative skewness and positive excess kurtosis,
see Figure~\ref{fig4.1b.0}, 
and the impact of the fourth cumulant remains minimal.

\begin{figure}[H]
\hskip1.42cm
\includegraphics[width=0.80516\textwidth,height=6cm]{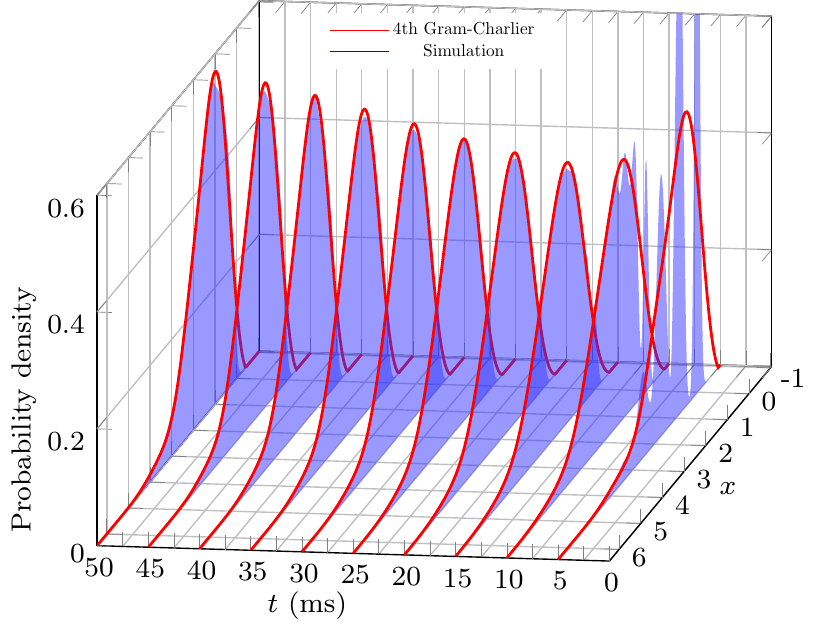} 
\vskip-.2cm
\caption{\small Fourth-order Gram-Charlier expansions vs simulated densities.} 
\label{fig4.1a}
\end{figure}

\vspace{-0.3cm}

\noindent
Figure~\ref{fig4.1a} presents time-dependent fourth-order
 Gram-Charlier expansions \eqref{gram_charlier}, 
 based on exact moment formulas  
 at different times for the probability density function of $V_{\scaleto{\rm I}{4.4pt}} (t )$. 

 \medskip

\begin{figure}[H]
\centering
\begin{subfigure}{.5\textwidth}
\centering
\includegraphics[width=1\textwidth]{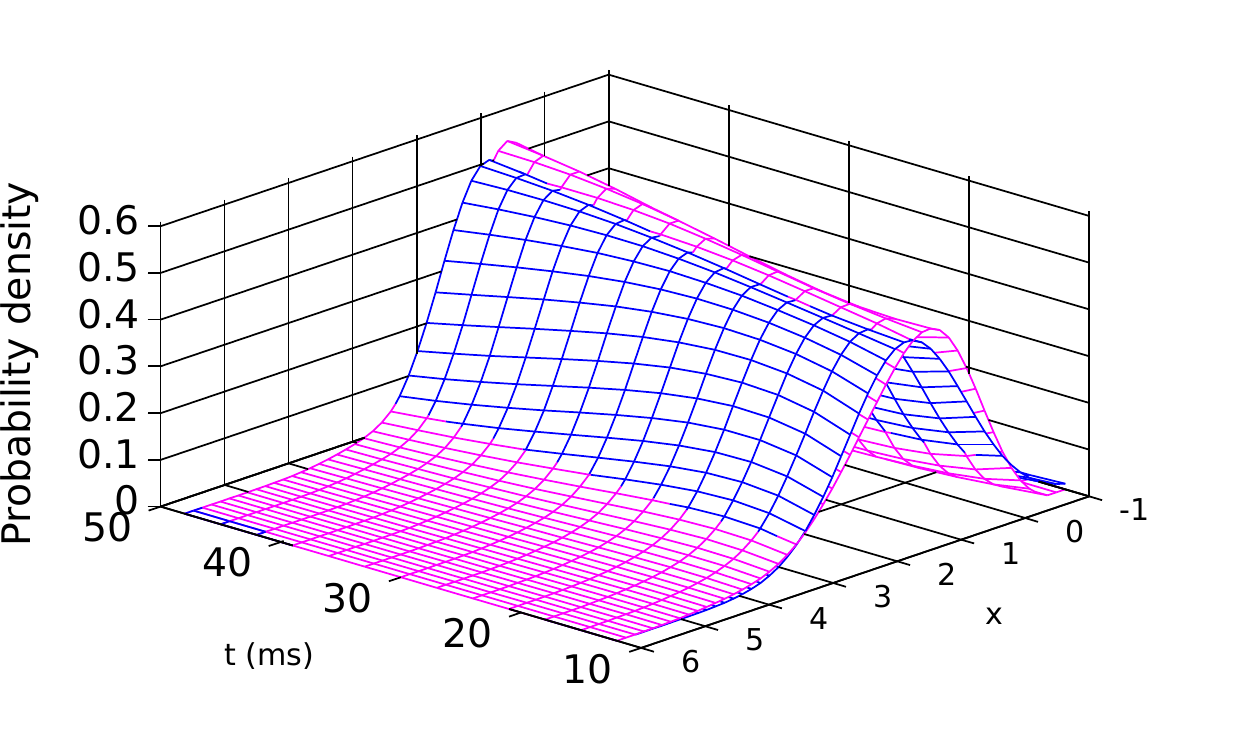} 
\caption{\small Gaussian diffusion {\em vs} $4^{\rm th}$ Gram-Charlier.} 
\end{subfigure}
\hskip-0.2cm
\begin{subfigure}{.5\textwidth}
\centering
\includegraphics[width=1\textwidth]{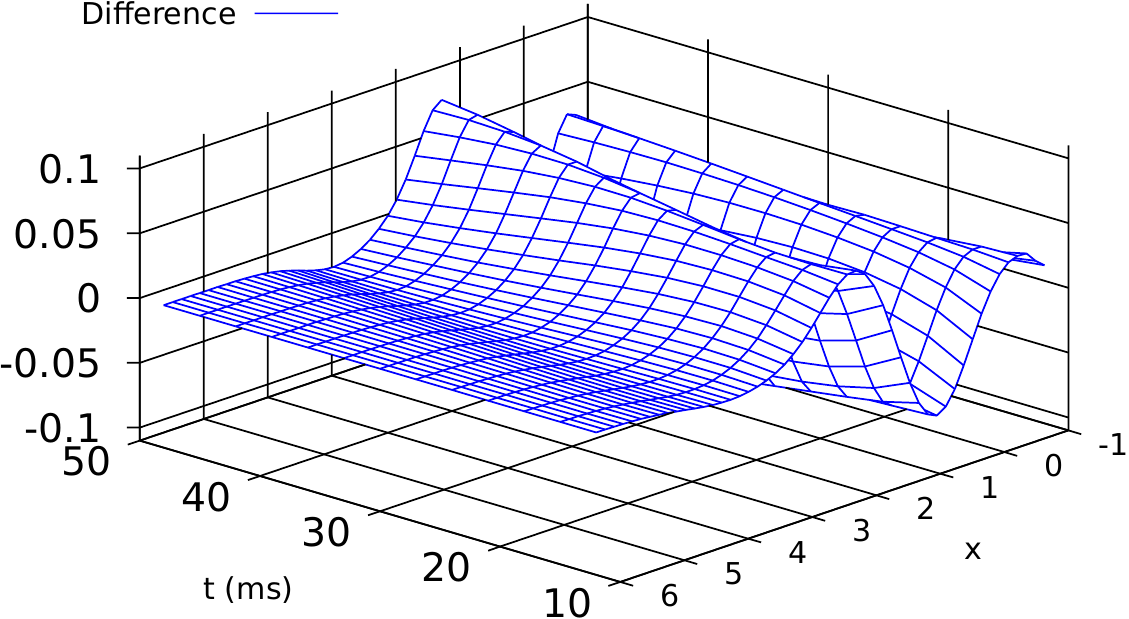} 
\vskip0.35cm
\caption{\small Difference between $2^{\rm nd}$ and $4^{\rm th}$ expansions.} 
\end{subfigure}
\caption{\small Fourth-order Gram-Charlier expansion {\em vs} diffusion approximation.} 
\label{fig6}
\end{figure}

 As can be checked from Figure~\ref{fig4.1a}, 
 the fourth-order Gram-Charlier expansions fit the purple areas obtained by 
 Monte Carlo simulations. 
 Figure~\ref{fig6}-$a)$ 
 compares the Gaussian diffusion
 (blue) approximation to the fourth-order Gram-Charlier expansion 
 (purple) for the probability density function of $V_{\scaleto{\rm I}{4.4pt}} (t )$ 
 while Figure~\ref{fig6}-$b)$ 
 represents the relative
 difference between the Gaussian diffusion and fourth-order approximations. 
 
\vskip0.2cm

\vspace{-0.6cm}

\section*{Conclusion}
This paper presents closed-form expressions for the cumulants of arbitrary orders
of filtered multivariate Hawkes processes with excitation and inhibition, for
application to the modeling of spike trains.
Such expressions can be used for the prediction and sensitivity analysis
of the statistical behavior of the model over time
via immediate numerical evaluations over multiple ranges of parameters,
whereas Monte Carlo estimations appear slower and less reliable.
They are also used to estimate the 
probability densities of neuronal membrane potentials
using Gram-Charlier density expansions.

\appendix

\section{Proofs of joint cumulant identities} 
\label{s7}
\noindent 
 In this section we extend the algorithm of \cite{hawkescumulants,hawkescumulantsc}
 for the recursive calculation of the joint cumulants of
 of Hawkes point processes from the univariate to multivariate setting. 
 We consider a self-exciting point process on $\X:=(\real^d) \times \{1,\ldots , m\}$,
 $d\geq 1$, with Poisson offspring intensities
 $\gamma_i (dx \times \{i\}) = \gamma_{i,j}(dx)= \gamma_{i,j}(x)dx$ and Poisson
 baseline intensity $\nu (dx \times \{j\}) = \nu_j (dx)= \nu_j (x)dx$
 on each copy of $\real^d$, $j=1,\ldots , m$.
 This process is built
 in the cluster process framework of \cite{hawkes} 
 on the space
$$
 \Omega = \big\{
 \xi = \{ (x,i) \}_{i\in I} \subset \X \ : \
 \#( A \cap \xi ) < \infty 
 \mbox{ for all compact } A\in {\cal B} ( \X ) 
 \big\}
 $$
 of locally finite configurations on $\X$, whose elements 
 $\xi \in \Omega$ are identified with the Radon point measures 
 $\displaystyle \xi (dz) = \sum_{(x,i)\in \xi} \epsilon_{(x,i)} (dz)$, 
 where $\epsilon_{(x,i)}$ denotes the Dirac measure at $(x,i)\in \X$. 
 Any initial point $(x,i) \in \X$ 
 branches into a Poisson random sample on $\X$, denoted by 
 $\xi_{\gamma_i} (d z )$, 
 with intensity measure   
 $\gamma_{i,j} (y+dx )$ on every copy of $\real^d$,
 $i,j=1,\ldots , m$.
 In case $d=1$ and $\gamma_{i,j}(s) = 0$ for $s\leq 0$, 
$$
   \displaystyle N_t^{(i)} (\xi ) := \xi ( [0,t] \times \{i\})
   = \sum_{x \ \! : \ \! (x,i) \in \xi} {\bf 1}_{[0,t]}(x),
   \qquad
   i = 1, \ldots , m,
$$ 
 represents a multivariate Hawkes process with stochastic intensities
 of the form 
$$ 
 \lambda^{(i)}_t := \nu + \sum_{j=1}^m \int_0^t \gamma_{i,j} (t-s) dN^{(j)}_s,
 \qquad t\in \real_+,
 \quad i = 1, \ldots , m. 
$$
 For $f$ a sufficiently integrable real-valued function on $\X$, we let
$$ 
 G_{(x,i)} (f) = f (x,i) \E_i \left[ \prod_{(y,j)\in \xi} f (y+x,j) \right] 
$$ 
 denote the Probability Generating Functional (PGFl) of the branching process
 $\xi$ given that it is started from a single point at $(x,i) \in \X$. 
The next proposition states a recursive property
for the Probability Generating Functional $G_{(x,i)}(f)$, see
also Theorem~1 in \cite{adamopoulos}. 
 \begin{prop}
  \label{fjklds}
  The Probability Generating Functional $G_{(x,i)}(f)$ satisfies
$$ 
 G_{(x,i)} (f) 
 = f(x,i) \exp \left( \sum_{j=1}^m
\int_{\real^d} ( G_{(x+y,j)} ( f ) - 1 ) \gamma_{i,j} ( dy) \right),
 \qquad (x,i)\in \X, 
$$ 
 and the PGFl of the Hawkes process with Poisson baseline intensity
 $\nu$ on $\X$ is given by
$$ 
   G_\nu (f)
   =
   \exp \left( \sum_{i=1}^m \int_{\real^d} ( G_{(x,i)} (f) - 1 ) \nu_i ( x ) dx \right). 
$$
\end{prop}
\begin{Proof} 
 Viewing the self-exciting point process $\xi$ as a marked
 point process we have, see e.g. Lemma~6.4.VI of \cite{daley}, 
\begin{eqnarray*} 
  \lefteqn{
    G_{(x,i)} (f) 
  =  f(x,i) \E_i\left[ \prod_{(y,j)\in \xi} f(y+x,j) \right] 
  }
  \\
 & = & f(x,i) \E_i \left[ \prod_{(y,j)\in \xi_{\gamma_j}} \left( \prod_{(z,k)\in \xi} f(z+y +x, k ) \right) \right] 
\\
 & = & f(x,i) \E_i \left[ \prod_{(y,j)\in \xi_{\gamma_j}} \E_j \left[ \prod_{(z,k)\in \xi} f(x+y+z,k) \right] \right] 
\\
 & = & f(x,i) \E_i \left[ \prod_{(y,j)\in \xi_{\gamma_j}} G_{(x+y,j)} ( f ) \right] 
\\
& = & \re^{-\gamma_i (\X) } f(x,i) \sum_{n=0}^\infty \frac{1}{n!}
\sum_{j_1,\ldots , j_n=1}^m
\int_{(\real^d)^n} G_{(x+y_1,j_1)}(f) \cdots G_{(x+y_n,j_n)} (f) \gamma_{i,j_1} (dy_1)\cdots \gamma_{i,j_n} ( dy_n) 
\\
& = & f(x,i) \exp \left( \sum_{j=1}^m
\int_{\real^d} ( G_{(x+y,j)} ( f ) - 1 ) \gamma_{i,j} ( dy) \right)
, 
\end{eqnarray*} 
 and 
\begin{eqnarray*} 
  G_\nu (f)
  & = & 
  \re^{ - \nu ( \X )
  } \sum_{n=0}^\infty \frac{1}{n!}
  \sum_{j_1,\ldots , j_n=1}^m
  \int_{ (\real^d)^n } G_{(y_1,j_1)}(f) \cdots G_{(y_n,j_n)}(f) \nu_{j_1} ( dy_1 ) \cdots \nu_{j_n} ( dy_n ) 
  \\
    & = &
  \exp \left( \sum_{i=1}^m \int_{\real^d} ( G_{(x,i)} (f) - 1 ) \nu_i ( x )dx \right). 
\end{eqnarray*} 
\end{Proof}
\noindent
 Let
$$ 
 M_{(x,i)} (f) := G_{(x,i)} \big(e^f\big) = \E_i \left[ \exp \left( f(x,i) +
 \sum_{(y,j)\in \xi} f(x+y,j) \right) \right] 
$$ 
 denote the Moment Generating Functional (MGFl) of the
 random sum $\displaystyle \sum_{(y,j)\in \xi} f(y,j)$ 
 given that the cluster process $\xi$ starts 
 from a single point at $(x,i)\in \X$. 
 The following corollary is an immediate consequence of
 Proposition~\ref{fjklds}, see also Proposition~2.6 in
 \cite{bogachev2} for Poisson cluster processes.
\begin{corollary} 
  The Moment Generating Functional $M_{(x,i)}(f)$ satisfies
  the recursive relation 
\begin{equation}
   \label{mfdsf0} 
 M_{(x,i)} (f) 
 = \exp \left( f(x,i) + \sum_{j=1}^m \int_{\real^d} ( M_{(x+y,j)} ( f ) - 1 ) \gamma_{i,j} ( dy) \right), 
 \qquad (x,i)\in \X.
\end{equation} 
 The MGFl of the Hawkes process with baseline intensity $\nu$ on $\X$ is given by
 \begin{equation}
   \label{mfdsf} 
 M_\nu (f)
 = 
 \exp \left( \sum_{i=1}^m \int_{\real^d} ( M_{(x,i)} (f) - 1 ) \nu_i ( x )dx \right). 
\end{equation} 
\end{corollary}
\begin{Proofy} {\em of Proposition~\ref{p1}.} 
  For simplicity, the proof is written
  using Bell polynomials in the univariate case
  for $\kappa_{(x,i)}^{(n)}(f): = \kappa_{(x,i)}^{(n)}(f,\ldots , f)$
  with $f = f_1 = \cdots = f_n$,
  and the general case is deduced by polarization. 
  By \eqref{mfdsf0}, \eqref{cgf} and the Fa\`a di Bruno formula  \eqref{dfjkl0}, 
  we have
 \begin{align} 
   \nonumber 
 & 
     \sum_{n=1}^\infty \frac{t^n}{n!} \kappa_{(x,i)}^{(n)}(f)
 = 
  \log M_{(x,i)} ( t f) 
     \\
   \nonumber 
   & =  tf(x,i) + \sum_{j=1}^m \int_{\real^d}
   \big( \re^{\log M_{(x+y,j)} ( tf )} - 1 \big) \gamma_{i,j} ( dy)
\\
\label{fjlkdsf} 
 & =  tf(x,i)
 +
 t 
 \sum_{j=1}^m
 \int_{\real^d}
 \kappa_{(x+y,j)}^{(1)}(f)
 \gamma_{i,j} ( dy)
 +
 \sum_{n=2}^\infty
 \frac{t^n}{n!}
 \sum_{j=1}^m
 \int_{\real^d}
 B_n \big( \kappa_{(x+y,j)}^{(1)}(f), \ldots , \kappa_{(x+y,j)}^{(n)}(f) \big)
 \gamma_{i,j} ( dy). 
\end{align} 
 At the first order, the expansion \eqref{fjlkdsf} yields 
\begin{eqnarray*} 
   \kappa_{(x,i)}^{(1)}(f) & = & f(x,i) + \int_{\real^d} \sum_{j=1}^m
   \kappa_{(x+y,j)}^{(1)}(f) \gamma_{i,j} ( dy)
\\
  & = & f(x,i) + \sum_{n=1}^\infty
  \sum_{j_1,\ldots , j_n=1}^m
  \int_{\real^d} \cdots \int_{\real^d} f(x+y_1+\cdots + y_n,j_n)
  \gamma_{i,j_2} ( dy_1) \cdots \gamma_{j_{n-1},j_n} ( dy_n), 
\end{eqnarray*} 
 while at the order $n\geq 2$ it shows that 
\begin{eqnarray*} 
 \kappa_{(x,i)}^{(n)}(f) & = &  
 \sum_{j=1}^m
 \int_{\real^d}
  B_n \big( \kappa_{(x+y,i)}^{(1)}(f), \ldots , \kappa_{(x+y,i)}^{(n)}(f) \big)
  \gamma_{i,j} ( dy)
  \\
   & = &  
 \big( \Gamma \kappa_{(\cdot ,\cdot )}^{(n)}(f) \big) (x,i)
 +
 \big(
 \Gamma 
 \big( B_n \big( \kappa_{(\cdot ,\cdot )}^{(1)}(f), \ldots , \kappa_{(\cdot ,\cdot )}^{(n)}(f) \big)
 - \kappa_{(\cdot ,\cdot )}^{(n)}(f) \big)
 \big) (x,i). 
\end{eqnarray*} 
 The above relation rewrites as 
 $$ 
 \big( (I - \Gamma ) \kappa_{(\cdot ,\cdot )}^{(n)}(f) \big) (x,i)
  = 
 \Gamma 
 \big( B_n \big( \kappa_{(\cdot ,\cdot )}^{(1)}(f), \ldots , \kappa_{(\cdot ,\cdot )}^{(n)}(f) \big)
 - \kappa_{(\cdot , \cdot )}^{(n)}(f) \big) (x,i), 
$$ 
 which yields 
\begin{align*} 
 & 
 \kappa_{(x,i)}^{(n)}(f) 
 = 
 \big(
 ( I - \Gamma )^{-1} \Gamma 
 \big( B_n \big( \kappa_{(\cdot , \cdot )}^{(1)}(f), \ldots , \kappa_{(\cdot , \cdot )}^{(n)}(f) \big)
 - \kappa_{(\cdot , \cdot )}^{(n)} (f) \big)
 \big) (x,i) 
 \\
 & =  
 \sum_{p=1}^\infty \sum_{i_1,\ldots , i_p=1}^m 
 \int_{\real^d} \cdots \int_{\real^d}
  \\
 & \quad
\big( B_n \big( \kappa_{(x+x_1+\cdots + x_p,i_p)}^{(1)}(f), \ldots , \kappa_{(x+x_1+\cdots + x_p,i_p)}^{(n)} (f)\big)
 - \kappa_{(x+x_1+\cdots + x_p,i_p)}^{(n)} (f)\big)
 \gamma_{i,i_1} ( dx_1) \cdots \gamma_{i_{p-1},i_p} ( dx_p ) 
\\
 & =  
 \sum_{p=1}^\infty
 \sum_{i_1,\ldots , i_p=1}^m
 \sum_{k=2}^n
 \\
 & \quad
  \int_{\real^d} \cdots \int_{\real^d} 
  B_{n,k} \big( \kappa_{(x+x_1+\cdots + x_p,i_p)}^{(1)}(f), \ldots , \kappa_{(x+x_1+\cdots + x_p,i_p)}^{(n-k+1)}(f) \big)
 \gamma_{i,i_1} ( dx_1) \cdots \gamma_{i_{p-1},i_p} ( dx_p ), 
\end{align*} 
 $n \geq 2$.
\end{Proofy}

\begin{Proofy} {\em of Corollary~\ref{c1}.} 
\noindent
As above, the proof is only written using Bell polynomials
in the case $f = f_1= \cdots = f_n$. 
 By \eqref{cgf}, \eqref{mfdsf}
 and the Fa\`a di Bruno formula \eqref{dfjkl0}, we have
\begin{eqnarray*} 
 \sum_{n=1}^\infty \frac{t^n}{n!} \kappa^{(n)}(f)
& = &
 \log M_\nu (tf)
 \\
 & = &  
 \sum_{i=1}^m \int_{\real^d} ( M_{(x,i)} (tf) - 1 ) \nu_i ( x )dx
 \\
  & = &  
 \sum_{i=1}^m \int_{\real^d} ( \re^{\log M_{(x,i)} (tf)} - 1 ) \nu_i ( x )dx
  \\
  & = &  
  \sum_{i=1}^m
  \sum_{n=1}^\infty
 \frac{t^n}{n!}
 B_n \big( \kappa_{(x,i)}^{(1)}(f) , \ldots , \kappa_{(x,i)}^{(n)}(f) \big)
 \nu_i ( x)dx, 
\end{eqnarray*}
and therefore
$$ 
\kappa^{(n)} (f) =
\sum_{i=1}^m
\int_{\real^d}
B_n \big( \kappa_{(x,i)}^{(1)}(f) , \ldots , \kappa_{(x,i)}^{(n)}(f) \big) 
 \nu_i ( x)dx, \qquad n \geq 2.
$$ 
\end{Proofy} 

\begin{Proofy} {\em of Lemma~\ref{l1}.} 
 Here we take $d=1$. For all $p,\eta \geq 0$ we have the equalities 
     \begin{eqnarray*} 
       \lefteqn{
         ( ( I - \Gamma )^{-1} \Gamma 
 e_{p,\eta, t,k} ) ( x,i ) 
       }
       \\
        & = & 
 \sum_{n=1}^\infty
 \sum_{j_1,\ldots , j_{n-1}=1}^m
 \int_{[0,t]^n} 
 e_{p,\eta, t,k} ( x+x_1+\cdots + x_n )
 \gamma_{i,j_1} ( dx_1 ) \cdots \gamma_{j_{n-1},k} ( dx_n )  
\\
 & = & 
 \sum_{n=1}^\infty
\frac{[W^n]_{i,k}}{(n-1)!}
 \int_0^{t-x}
 (x+y)^p
 \re^{\eta (x+y)}
 y^{n-1} \re^{-by}
 dy
\\
 & = & 
 \re^{\eta x} \int_0^{t-x}
 (x+y)^p [ W \re^{y W}]_{i,k} 
 \re^{(\eta -b)y}
 dy
 \\
  & = & 
 \int_0^{t-x}
 e_{p,\eta, t,k} (x+y) 
 [ W \re^{y W}]_{i,k} 
 \re^{-by}
 dy, \quad x\in [0,t],  
\end{eqnarray*} 
     where we used the fact that the sum $\tau_1+\cdots +\tau_n$
     of $n$ exponential random variables with parameter $b>0$ has
     a gamma distribution with shape parameter $n\geq 1$ and scaling parameter
     $b>0$.
\end{Proofy}
\section{Computer codes} 
\label{a2}
 \noindent 
 The recursion \eqref{fjkl} and Equation~\eqref{al} can be implemented
 for any family $(g_1,\ldots , g_n)$ of functions defined on $\real_+$
 in the following Maple code. 
 The joint cumulants 
 $\langle \langle V_{l_1}(t_1 ) \cdots V_{l_n}(t_n ) \rangle\rangle$,
 are obtained for $1 \leq l_1,\ldots , l_n \leq m$ using the command
 $\verb|c(W,b,{g1,...,gn},{l1,...,ln},|\\ \verb|{t1,...,tn},mu)|$ 
 in the code below. 

 \medskip

\begin{lstlisting}[language=Maple]
with(LinearAlgebra):
a := proc(y) option remember; return evalf(Multiply(W, MatrixExponential(W, y))); end proc;
h := proc(z, j, W, b, g::list, l::list, t::list) local p, q, r, s, y, i, m, n, k, c; option remember; n := nops(t); if n = 1 then return evalf(g[1](z, t[1])*charfcn[j](l[1]) + int(g[1](z + y, t[1])*exp(-b*y)*a(y)[j, l[1]], y = 0 .. t[1] - z)); end if; s := 0; r := Iterator:-SetPartitions(n); for q in r do p := r:-ToSets(q); if 2 <= nops(p) then for k to Dimension(W)[1] do c := exp(-b*y)*a(y)[j, k]; for i to nops(p) do c := c*h(z + y, k, W, b, map(op, convert(p[i], list), g), map(op, convert(p[i], list), l), map(op, convert(p[i], list), t)); end do; s := s + c; end do; end if; end do; return int(s, y = 0 .. t[1] - z); end proc;
c := proc(W, b, g::list, l::list, t::list, mu::list) local y, e, p, q, r, s, i, j, m, n; option remember; n := nops(t); s := 0; for j to Dimension(W)[1] do s := s + mu[j](y)*h(y, j, W, b, g, l, t); if 2 <= n then r := Iterator:-SetPartitions(n); for q in r do p := r:-ToSets(q); if 2 <= nops(p) then e := 1; for i to nops(p) do e := e*h(y, j, W, b, map(op, convert(p[i], list), g), map(op, convert(p[i], list), l), map(op, convert(p[i], list), t)); end do; s := s + mu[j](y)*e; end if; end do; end if; end do; return int(s, y = 0 .. t[1]); end proc;
\end{lstlisting} 

\vspace{-0.7cm}

\noindent
 Joint moments 
 can be computed in Maple using the command
 $\verb|m(W,b,{g1,....,gn},{jl,|$\\$\verb|...,jl},{t1,...,tn},mu)|$ defined in the
 following code. 

\medskip

\begin{lstlisting}[language=Maple]
m := proc(W, b, f::list, l::list, t::list, mu::list) local e, u, p, q, r, s, i, n; option remember; n := nops(t); s := c(W, b, f, l, t, mu); if 2 <= n then r := Iterator:-SetPartitions(n); for q in r do p := r:-ToSets(q); if 2 <= nops(p) then e := 1; for i to nops(p) do e := e*c(W, b, map(op, convert(p[i], list), f), map(op, convert(p[i], list), l), map(op, convert(p[i], list), t), mu); end do; s := s + e; end if; end do; end if; return s; end proc;  
\end{lstlisting}

\vskip-0.7cm

\noindent
 Alternatively, the computation of joint cumulants can be carried out using
 the command $\verb|c[W,b,[g1,...,gn],[l1,...,ln],[t1,...,tn],mu]|$ 
 in the following Mathematica codes.
 
\medskip
 
\begin{lstlisting}[language=Mathematica]
Needs["Combinatorica`"]
a[y_] := W . MatrixExp[y*W];
h[z_, j_Integer, W_, b_, g__, l__, t__] := h[z, j, W, b, g, l, t] = (Module[{y, k, i, c, n, m, s}, n = Length[t]; If[n == 1, Return[g[[1]][z, t[[1]]]*Boole[j == l[[1]]] + Integrate[g[[1]][z + y, t[[1]]]*E^(-b*y)*a[y][[j, l[[1]]]], {y, 0, t[[1]] - z}]]]; s = 0; Do[c = 1; If[Length[p] >= 2, For[i = 1, i <= Length[p], i++, c *= Block[{u = y + z, w = g[[p[[i]]]], r = l[[p[[i]]]], v = t[[p[[i]]]]}, h[u, k, W, b, w, r, v]]]; s += c], {p, SetPartitions[n]}]; Return[Sum[Integrate[E^(-b*y)*a[y][[j, k]]*s, {y, 0, t[[1]] - z}], {k, 1, Dimensions[W][[1]]}]]]);
c[W_, b_, g__, l__, t__, mu_] := (Module[{y, e, n, i, j, m, s}, n = Length[g]; s = 0; For[j = 1, j <= Dimensions[W][[1]], j++, Do[e = mu[y][[j]]; For[i = 1, i <= Length[p], i++, e *= Block[{u = y, w = g[[p[[i]]]], r = l[[p[[i]]]], v = t[[p[[i]]]]}, h[u, j, W, b, w, r, v]]]; s += Flatten[{e}][[1]], {p, SetPartitions[n]}]]; Return[Integrate[s, {y, 0, t[[1]]}]]]);
\end{lstlisting}

\vspace{-0.7cm} 

\noindent 
Figures~\ref{fig4.11-0} to \ref{fig4.22-0} can also been plotted from 
the following Mathematica commands.\footnote{Mathematica computation times are significantly higher, probably due to the way recursions are carried out.} 

\begin{table}[H] 
  \centering
\scriptsize 
  \begin{tabular}{|l|l|c|}
 \hline
 \multicolumn{3}{|l|}{
  W := $\{\{$10, 0, 10, 0$\}$, $\{$ 0, 10, 10, -8 $\}$, $\{$ 10, 10, 0, -8 $\}$, $\{$ 10, 10, 10, -10$\}\}$;
}
\\
 \hline
 \multicolumn{3}{|l|}{
g[u\_, t\_] := E\string^(-(t - u)/0.01); {\rm mu}[t\_] := $\{$ 250, 250, 250, 250 $\}$;   
} 
\\
\hline
\hline
Instruction & Computed quantity & Computation time
 \\ 
 \hline
 \hline
c[W, 50, $\{$g$\}$, $\{$2$\}$, $\{$t$\}$, mu]  & First cumulant of V2(t)  & One second
\\
\hline
c[W, 50, $\{$g,g$\}$, $\{$4,4$\}$, $\{$t,t$\}$, mu] & Second cumulant of V4(t) & 122 seconds
\\
\hline
c[W, 50, $\{$g,g$\}$, $\{$4,2$\}$, $\{$t,0.05$\}$, mu] & Covariance of (V2(t1),V4(t)) for t<t1=0.05 & 250 seconds
\\
\hline
c[W, 50, $\{$g,g$\}$, $\{$2,4$\}$, $\{$0.05,t$\}$, mu] & Covariance of (V2(t1),V4(t)) for t>t1=0.05 & 250 seconds 
\\
\hline
c[W, 50, $\{$g,g,g$\}$, $\{$4,4,4$\}$, $\{$t,t,t$\}$, mu] & Third cumulant of V4(t) & 3057 seconds 
\\
\hline
c[W, 50, $\{$g,g,g$\}$, $\{$4,1,1$\}$, $\{$t,0.05,0.05$\}$, mu] & Third joint cumulant of (V1(t1),V1(t1),V4(t)), t<t1=0.05 & \multicolumn{1}{c}{} 
\\
\cline{1-2}
c[W, 50, $\{$g,g,g$\}$, $\{$1,1,4$\}$, $\{$0.05,0.05,t$\}$, mu] & Third joint cumulant of (V1(t1),V1(t1),V4(t1)), t>t1=0.05 &  \multicolumn{1}{c}{} 
\\
\cline{1-2}
c[W, 50, $\{$g,g,g,g$\}$, $\{$4,4,4,4$\}$, $\{$t,t,t,t$\}$, mu]  & Fourth cumulant of V4(t) &  \multicolumn{1}{c}{} 
\\
\cline{1-2}
c[W, 50, $\{$g,g,g,g$\}$, $\{$1,2,3,4$\}$, $\{$t,t,t,t$\}$, mu]  & Fourth joint cumulant of (V1(t),V2(t),V3(t),V4(t)) & \multicolumn{1}{c}{}  
\\ 
\cline{1-2}
\end{tabular}
\end{table} 

\vskip-0.3cm

\noindent
 Standard moments of order $n\geq 1$ can be computed in Mathematica using the command
 $\verb|m[W,b,[g1,...,gn],[l1,...,ln],[t1,...,tn],mu]|$
 defined below.

\medskip

\begin{lstlisting}[language=Mathematica]
m[W_, b_, g__, l__, t__, mu_] := (Module[{n, e, i, s}, s = 0; n = Length[t]; If[n == 0, Return[1]]; Do[e = 1; For[i = 1, i <= Length[pp], i++, e *= c[W, b, g[[pp[[i]]]], l[[pp[[i]]]], t[[pp[[i]]]], mu]]; s += e, {pp, SetPartitions[n]}]; Flatten[{s}][[1]]]);
\end{lstlisting}

\vspace{-0.7cm} 

\section{Joint cumulants and Fa\`a di Bruno formula} 
We refer to e.g. \cite{elukacs} or \cite{mccullagh}
for the background combinatorics recalled in this section. 
The joint {cumulants} of orders $(l_1,\ldots , l_n)$
of a random vector $X=(X_1,\ldots , X_n)$, $1 \leq l_1,\ldots , l_n \leq m$, 
are the coefficients $\langle \langle X_1^{l_1} \cdots X_n^{l_n} \rangle \rangle$ 
 appearing in the log-moment generating (MGF) expansion 
\begin{equation} 
\label{cgf} 
\log \langle \re^{t_1X_1+\cdots + t_n X_n}\rangle 
 = 
 \sum_{l_1,\ldots , l_n\geq 1} \frac{t^{l_1}_1\cdots t^{l_n}_n}{l_1! \cdots l_n!}
 \langle \langle X_1^{l_1} \cdots X_n^{l_n} \rangle \rangle, 
\end{equation} 
for $(t_1,\ldots , t_n)$ in a neighborhood of zero in $\real^n$.
Recall that if $f(t)$ admits the formal series expansion
$$
f(t) = \sum_{n=1}^\infty \frac{a_n}{n!}t^n,
$$ 
by the Fa\`a di Bruno formula we have
\begin{equation} 
\label{dfjkl0} 
\re^{f ( t )} - 1 
 = 
 \sum_{n=1}^\infty
 \frac{t^n}{n!}
 B_n ( a_1, \ldots ,a_n ), 
\end{equation}
 where 
$$ 
 B_n ( a_1 , \ldots , a_n ) 
  = 
 \sum_{k=1}^n
 B_{n,k} ( a_1 , \ldots , a_{n-k+1} )
$$ 
 is the complete Bell polynomial of degree $n \geq 1$, 
 and
 $$ 
 B_{n,k} ( a_1 , \ldots , a_{n-k+1} ) 
=
 \sum_{\pi_1 \cup \cdots \cup \pi_k = \{ 1, \ldots , n \}} 
 a_{|\pi_1|}(X) \cdots a_{|\pi_k|}(X), 
 \qquad 1\leq k \leq n, 
$$ 
 is the partial Bell polynomial of order $(n,k)$,
 where the sum runs over the partitions 
 $\pi_1,\ldots , \pi_k$ of the set $\{ 1 , \ldots , n \}$,
 and $|\pi_i|$ denotes the cardinality of $\pi_i$. 
 Joint moments can be obtained from the joint moment-cumulant relation 
\begin{equation}
   \label{mcr} 
 \langle V_{l_1}(t_1 ) \cdots V_{l_n}(t_n ) \rangle 
= 
 \sum_{\pi \in \Pi [n]} 
 \prod_{j=1}^{|\pi|} 
 \left< \left< \prod_{i \in \pi_j }
 V_{l_i}(t_i ) 
 \right> \right> 
, 
\end{equation} 
  where the above sum is over the set 
$\Pi [n]$ of partitions $\pi$ of
$\{1,\ldots , n\}$. 
 Joint cumulants can also be recovered from joint moments
 from the relation 
\begin{equation} 
\nonumber 
 \langle \langle V_{l_1}(t_1 ) \cdots V_{l_n}(t_n) \rangle \rangle 
 = 
 \sum_{\pi \in \Pi [n]}   
 ({|\pi|}-1)! (-1)^{{|\pi|}-1} 
 \prod_{j=1}^{|\pi|} \left< \prod_{i \in \pi_j} V_{l_i}(t_i ) \right>, 
\end{equation} 
where the above sum is over the set 
$\Pi [n]$ of partitions $\pi$ of
$\{1,\ldots , n\}$, which 
can be obtained by M\"obius inversion of the moment-cumulant relation 
\eqref{mcr}.
 
\footnotesize 

\def\cprime{$'$} \def\polhk#1{\setbox0=\hbox{#1}{\ooalign{\hidewidth
  \lower1.5ex\hbox{`}\hidewidth\crcr\unhbox0}}}
  \def\polhk#1{\setbox0=\hbox{#1}{\ooalign{\hidewidth
  \lower1.5ex\hbox{`}\hidewidth\crcr\unhbox0}}} \def\cprime{$'$}

\end{document}